\documentclass[twocolumn,preprintnumbers,amsmath,amssymb,APSl,prd,nofootinbib,superscriptaddress]{revtex4}
\usepackage{amsmath, amsfonts, amsthm, amssymb, graphicx}
\usepackage{epstopdf}
 \usepackage{slashed}
 \usepackage{graphicx,epsfig}
\usepackage{amsmath}
\usepackage {amssymb}
\usepackage{color}          
\usepackage[utf8]{inputenc}
\usepackage{hyperref}
\usepackage[francais]{babel}
\usepackage[T1]{fontenc}
\usepackage{lmodern}
\allowdisplaybreaks

\newcommand{\dd}{\mathrm{d}}
\newcommand{\rhOh}{r_H^2 \Omega_H^2}
\newcommand{\rhoverr}{\left(\frac{r_H}{r}\right)}

\newcommand{\be}{\begin{eqnarray}}
\newcommand{\ee}{\end{eqnarray}}
\newcommand{\bea}{\begin{eqnarray}}
\newcommand{\eea}{\end{eqnarray}}

\makeatletter
\def\@appendixcntformat#1{\csname the#1\endcsname}
\def\@hangfrom@appendix#1#2#3{%
  #1#2\@if@empty{#3}{}{\quad#3}%
}
\makeatother

\begin{document}

\title{Neural-Spectral Discovery of Rotating Black Holes Beyond General Relativity}

\author{Felipe Agurto-Sepúlveda}
\altaffiliation{These authors contributed equally to this work.}
\affiliation{Departamento de Física de Partículas $\&$ Instituto Galego de Física de Altas Enerxías (IGFAE),
Universidade de Santiago de Compostela, E-15782 Santiago de Compostela, Spain}

\author{Marcelo Oyarzo}
\altaffiliation{These authors contributed equally to this work.}
\affiliation{Departamento de Física de Partículas $\&$ Instituto Galego de Física de Altas Enerxías (IGFAE),
Universidade de Santiago de Compostela, E-15782 Santiago de Compostela, Spain}

\author{Anxo Biasi}
\affiliation{Departamento de Física de Partículas $\&$ Instituto Galego de Física de Altas Enerxías (IGFAE),
Universidade de Santiago de Compostela, E-15782 Santiago de Compostela, Spain}

\author{Devansh Agarwal}
\affiliation{West Virginia University, Department of Physics and Astronomy, P. O. Box 6315, Morgantown, WV, USA}
\affiliation{Center for Gravitational Waves and Cosmology, West Virginia University, Chestnut Ridge Research Building, Morgantown, WV, USA}

\author{Ethan Tregidga}
\affiliation{Laboratoire d’Astrophysique, EPFL, Observatoire de Sauverny, 1290 Versoix, Switzerland}

\author{James F. Steiner}
\affiliation{AstroAI, Center for Astrophysics | Harvard $\&$ Smithsonian, 60 Garden Street, Cambridge, MA 02138, USA}

\author{Jos\'e D. Edelstein}
\affiliation{Departamento de Física de Partículas $\&$ Instituto Galego de Física de Altas Enerxías (IGFAE),
Universidade de Santiago de Compostela, E-15782 Santiago de Compostela, Spain}

\author{Gaston Giribet}
\affiliation{Center for Cosmology and Particle Physics Department of Physics, New York University, 726 Broadway, New York City, NY10003, USA}

\author{Cecilia Garraffo}
\thanks{Senior Author}
\affiliation{AstroAI, Center for Astrophysics | Harvard $\&$ Smithsonian, 60 Garden Street, Cambridge, MA 02138, USA}



\begin{abstract}
Finding rotating black hole solutions in higher-curvature theories of gravity is a problem of fundamental importance. Virtually every approach to reconcile gravity with quantum mechanics predicts corrections to the Einstein-Hilbert action, yet no systematic solution-generating method exists for the stationary sector. We close this gap with {\sc Akribeia}, a novel hybrid framework that pairs physics-informed neural networks with a pseudo-spectral refinement step, yielding certified neural-field rotating black hole solutions ---continuous, globally defined  functions, parametric in the coupling constants--- whose residuals against the field equations are verified to extreme precision. We apply the method to theories quadratic and cubic in the curvature and construct, for the first time, families of rotating black holes featuring multiple non-vanishing angular momenta, parametric in the new coupling constants. After validating against previously known five-dimensional spacetimes, we present new solutions in scenarios leading to a highly non-linear/non-perturbative coupled system of ordinary differential equations. Our method can be systematically adapted to other setups involving partial differential equations as well.
\end{abstract}

\maketitle

\section{Introduction}

Any sensible approach to quantum gravity induces higher curvature corrections to the gravitational action. Even in the semi-classical approach, in which the quantum field theory of matter is formulated on a classical spacetime, higher curvature terms are induced in the gravitational effective action. The role that higher curvature terms would play in the ultraviolet regime of gravitational theory was studied early on, and it was quickly understood that they would lead to the improvement of its seemingly bad behavior \cite{Stelle:1976gc}. Later, string theory showed how these terms arise naturally \cite{Gross:1986iv, Zwiebach:1985uq}.

The advent of precision strong-field gravity has transformed the search for departures from General Relativity (GR) into an empirical program, and this is also one of the motivations we have to investigate this problem: {The Event Horizon Telescope has resolved the horizon-scale emission of M87* and Sgr A*, enabling direct tests of the Kerr geometry \cite{EventHorizonTelescope:2019ggy, EventHorizonTelescope:2022xqj} (see also \cite{Vagnozzi:2022moj}), while the LIGO-Virgo-KAGRA network now catalogs over two hundred binary coalescences whose ringdown spectra probe the dynamical, highly curved regime near the merger \cite{LIGOScientific:2025rid}. Next-generation facilities are forecast to tighten constraints on deviations from Kerr by orders of magnitude \cite{Maselli:2023khq}. Because any such deviation would most naturally arise from higher-curvature corrections to the Einstein-Hilbert action, the construction and characterization of rotating black hole solutions in higher-curvature gravity is going to be relevant for confronting these theories with observation \cite{Cano:2024ezp}.} We hope that an adaptation of our analysis for the stationary case in 3+1 dimensions will be useful in this context.

A very useful model for studying the effects of higher curvature corrections is Lovelock theory \cite{Lanczos:1938sf, Lovelock:1971yv}. This is the most general gravitational theory with second-order field equations given by a symmetric and divergence-free rank-2 tensor. In five and six dimensions it reduces to the quadratic Einstein-Gauss-Bonnet (EGB) gravity, while in seven dimensions and higher it also includes cubic and higher terms. The action of the theory takes the form 
\begin{equation}
S = \int \dd^D x \sqrt{-g}\, \sum_{p=1}^{[D/2]} \frac{\alpha_p}{2^p} \,\,\delta^{\mu_1 \nu_1 \cdots \mu_p \nu_p}_{\rho_1 \sigma_1 \cdots \rho_p \sigma_p}\, \prod_{r=1}^{p}\, R^{\rho_r \sigma_r}_{\ \ \ \mu_r \nu_r} ~,
\label{action}
\end{equation}
where $\alpha_p$ are coupling constants of mass dimension $2p-D$ and $\delta^{\mu ...\nu}_{\rho ... \sigma  } $ is the generalized antisymmetric Kronecker delta. We set the convention $\alpha_1=(16\pi G)^{-1}$. The coupling $\alpha_0$ gives the bare cosmological constant, which we set to zero. The sum over $p$ turns out to be finite, as the terms of order $\mathcal{O}(R^n)$ with $n>D/2$ identically vanish.

Lovelock gravity has many advantages, which make it tractable as a possible consistent model with higher-curvature corrections. In particular, it is free of ghosts around maximally symmetric solutions \cite{Boulware:1985wk} and does not introduce massive degrees of freedom for the graviton \cite{Stelle:1977ry}. These properties allowed the study of several aspects of these theories, such as their role in the AdS/CFT correspondence \cite{Camanho:2009hu, Camanho:2013pda}, gravitational phase transition \cite{Camanho:2012da, Camanho:2013uda}, critical models in Anti-de Sitter (AdS) \cite{Deser:2011xc}, among many others. Schwarzschild-like black hole solutions were thoroughly investigated \cite{Garraffo:2008hu, Camanho:2011rj, Camanho:2013kfa}.

However, in spite of all the nice properties that Lovelock theory exhibits, there is an obvious limitation. The field equations derived from (\ref{action}) are substantially more complex than those of Einstein's theory, as they include higher powers of second derivatives of the metric tensor and the Kerr-Schild ansatz fails to generically lead to the stationary solution \cite{Anabalon:2009kq, Ett:2011fy}. Although in some systems the field equations of Lovelock theory can be solved analytically, this is only true for cases that exhibit very high symmetry. In fact, even the analytical stationary solution describing a spinning black hole in vacuum is unknown for this theory. We should recall that, even in General Relativity, it took 48 years for Kerr to find the axially symmetric rotating black hole solution \cite{Kerr:1963ud}.

Finding a stationary solution, even at the quadratic level, has been an open problem for a long time despite the many efforts to tackle it. In the last two decades, there have been partial advances in the literature. An exact rotating solution was found for a specific value of the coupling constant in 5 dimensions \cite{Anabalon:2009kq}, a numerical solution for a wider range of parameters was presented \cite{Brihaye:2008kh}, as well as approximate solutions to leading order in the angular momentum parameter \cite{Kim:2007iw}. However, all these achievements are partial, and finding an efficient method to solve the field equations in the stationary case remains an open problem.

From a numerical perspective, the primary obstacle is not the solution of the stationary boundary-value problem itself, but the construction of an initial approximation from which high-precision solvers then converge to the desired solution \cite{Newton1, Newton2}. Existing studies of rotating black holes in EGB(-dilaton) gravity \cite{Kleihaus:2012qz, Brihaye:2010wx, Brihaye:2008kh, Maselli:2015tta, Kleihaus:2015aje}, indeed, rely on Newton-type iterative schemes, which achieve excellent accuracy once initialized sufficiently close to the target branch. The construction of such initial seeds typically relies on problem-dependent structures, such as perturbative regimes, continuation procedures from special limits, or prior physical insight. In their absence, generating suitable initial seeds becomes a huge challenge, significantly limiting the discovery of new stationary solutions. Identifying deviations from GR in upcoming observations requires a systematic, parametric catalogue of rotating black hole solutions across beyond-GR theories. That is why, it is desirable to develop schemes with reduced or no dependence on accurate pre-existing seeds.

Physics-informed neural networks (PINNs) \cite{PINNs1} open a third road to constructing solutions to differential equations. Rather than producing values on a discrete grid, PINNs represent each unknown field as a neural field\!\!: a continuous, differentiable function parametrized by the weights of a neural network. The network is trained by minimizing the residuals of the field equations and boundary conditions at sampled collocation points, with automatic differentiation providing exact derivatives of the network output. This reformulates the boundary-value problem as a global optimization over function space rather than a step-by-step propagation across grid points. PINNs can identify solution branches starting from random initializations \cite{ZouWang}, without requiring perturbative seeds or continuation paths, and have been applied successfully across a range of non-linear problems \cite{Raissi2020, GomezSerrano1, GomezSerrano2}.

The trained neural field can represent a continuous family of solutions across parameter space, whereas classical methods must be re-run for each parameter setting. It is evaluable at arbitrary points in the domain, differentiable to any order and globally defined, properties that make it composable with other mathematical structure such as symmetry transformations and downstream learning over families of solutions.

In gravitational physics, however, PINN applications have been primarily exploratory \cite{Juan, Luna, Choptuik, Cornell, Pombo}, reflecting their main limitation\!\!: stiffness and moderate accuracy, typically reaching residuals of order $10^{-2}$ to $10^{-4}$. Conventional spectral and finite-difference solvers remain the methods of choice when high precision is required, but they need accurate pre-existing seeds, and constructing those seeds may become the limiting factor in their workflow.

In this Letter we present\!\!: (a) new rotating black hole solutions in Lovelock theory, and (b) {\sc Akribeia}, a general hybrid neural-spectral method to unlock rotating solutions beyond General Relativity. As a proof of concept, we focus on cubic Lovelock theory in odd dimensions, whose co-homogeneity-1 rotating solutions are governed by a highly non-linear, non-perturbative system of coupled ordinary differential equations. We obtain such solutions with unprecedented precision for arbitrary values of the gravitational couplings and equal angular momenta, and discuss the physical properties and the theoretical and phenomenological avenues ---including the possible construction of template catalogues for upcoming observations--- that this method opens.

\section{Spectrally Refined Neural Fields}

Let us start by introducing in this Section our discovery-refinement framework that produces certified neural-field solutions to non-linear boundary-value problems\!\!: continuous, globally defined functions whose residuals are verified below a quantified threshold. The framework constructs these solutions without perturbative seeds or parameter continuation procedures. A PINN first identifies an approximate continuous family of solutions across parameter space from random initialization. A pseudo-spectral solver with extended-precision arithmetic then refines it, reaching residuals far below machine precision for a given parameter configuration. This hybrid strategy combines the exploratory capabilities of machine-learning-based methods with the accuracy of established numerical techniques, providing a high-precision framework for studying rotating black holes in higher-curvature gravity.

To our knowledge, no machine-learning method has previously found a rotating black-hole solution in higher-curvature gravity. Prior PINN applications in gravitational physics have addressed perturbations of known backgrounds \cite{Juan,Cornell,Pombo} or time-dependent matter dynamics \cite{Choptuik}, not the finding of stationary background solutions themselves. Existing rotating black-hole solutions in, say, EGB gravity \cite{Brihaye:2010wx,Brihaye:2008kh,Kleihaus:2012qz,Maselli:2015tta,Kleihaus:2015aje} rely on Newton iteration from perturbative or continuation-based seeds. The combination of a PINN with a high-precision pseudo-spectral refiner has only recently been explored in other contexts \cite{ZouWang2}.

We will construct in what follows such a hybrid neural-spectral framework to obtain the first accurate spinning black holes to date in Lovelock gravity beyond EGB, focusing for concreteness on the cubic theory and using the quadratic case to certify its accuracy. While slowly rotating solutions in the cubic case are known \cite{Yue:2011et}, fully non-linear rotating black hole solutions in cubic or higher Lovelock gravity for arbitrary couplings and angular momenta have remained inaccessible \cite{Adair:2020vso}. 

To construct the rotating solutions we employed PINNs \cite{PINNs1}, parameterizing each of the metric functions in \eqref{BHsolution} by a deep network $\mathcal{N}(\mathbf{x};\mathbb{W})$, with $\mathbb{W}$ representing all the trainable parameters of the network (\textit{i.e.}, weights and biases). We will represent the architecture by the list $[n_1,n_2,\dots,n_L]$, with $n_i$ being the dimension of the $i$-th layer. This allows us to compute the equations on each point as $\mathcal{E}(\mathbf{x}) =\mathcal{E} \left[\mathcal{N}(\mathbf{x}),\nabla\mathcal{N}(\mathbf{x}),...\right]$. The network can then be trained by minimizing the loss function \eqref{Losssss}.
%
%

We implement exact (strong) boundary conditions by reparameterizing the network output, which can always be done on a single direction of the domain, which in our case corresponds to $\mathbf{x}=(x,\alpha)\in \mathbb{R}^2$. We impose the strong boundary condition $\mathcal{F}(\mathbf{x}_0)$ for the radial direction $x$ and define the function
\begin{equation}
\widehat{\mathcal{N}}(\mathbf{x})=\mathcal{F}(\mathbf{x}_0)+\left(1-e^{-(x-x_0)}\right)\mathcal{N}(\mathbf{x};\mathbb{W}) ~,
\end{equation}
where $\mathcal{N}(\mathbf{x}; \mathbb{W})$ is the raw network output. Consequently, the optimization process is reduced to minimizing the mean squared residual of the differential equations $\mathcal{E}$, evaluated over a set of $N$ collocation points $x_i$ distributed within the integration domain 
\begin{equation}
\mathcal{L}_{\mathrm{equations}}(\mathbb{W}) = \frac{1}{N}\sum_{x \in \mathcal{I}} |\mathcal{E}\left[\mathcal{N}(\mathbf{x}),\nabla\mathcal{N}(\mathbf{x}),...\right]|^2 ~.
\label{Losssss}
\end{equation}
Fundamentally, all required derivatives are computed exactly via automatic differentiation.
To solve the system of equations, we represent the unknown functions using fully connected neural networks with \textit{Swish} activation functions for the hidden layers, adapting the network depth to the complexity of the problem.

The training strategy for the neural networks proceeds in four main stages, encapsulated within a global phase of $10^4-10^5$ steps. Initially, we employ the \textit{Adam} optimizer \cite{Kingma:2014vow} with a learning rate warm-up, which starts at $10^{-6}$ and increases linearly to reach $10^{-3}$. The second stage consists of keeping the learning rate stable at $10^{-3}$ to broadly explore the highly non-linear solution landscape. Once a promising valley is identified, the third stage refines the search using an adaptive learning rate decay based on the validation loss to improve the precision of the solution. In the fourth stage, we apply the \textit{L-BFGS} algorithm \cite{LBFGS}, a quasi-Newton second-order optimizer, which accelerates convergence, allowing us to reduce the loss function to $10^{-5}$ and the equation residuals to $10^{-3}$.

To push the precision much further, the trained neural network is ultimately used to initialize a classic pseudo-spectral Chebyshev method equipped with a Gauss-Newton optimizer and extended-precision arithmetic. This part rapidly reduces the residuals far below machine double-precision. Once the desired accuracy is reached, the approximate solution is written in terms of Chebyshev polynomials, having a function that is a high-precision approximation of the solution. It allows us to compute the exact derivatives at any point, and consequently, the residuals.

The complete description of {\sc Akribeia}, our hybrid neural-spectral algorithm combining physics-informed neural networks with pseudo-spectral methods, is provided in the Appendix. This includes the architecture and training protocol of the PINN stage, the pseudo-spectral refinement procedure, the L-BFGS optimization and convergence diagnostics. As a quantitative precision benchmark that validates our solution, this discovery-refinement framework recovers the Myers-Perry rotating black hole \cite{Myers:1986un} in the GR limit, starting from random initializations with extreme accuracy. A comparison between the numerical and exact solutions in the $L^2$ norm yields a relative error of $10^{-85}$, achieved by employing 300-digit extended-precision arithmetic.

\section{Rotating black hole solution}

For generic stationary rotating black holes in odd dimensions $D=2n+1$, the symmetry is $\mathbb{R}_t\times U(1)^n$, while in the equal-angular-momenta sector it is enhanced to
$\mathbb{R}_t\times U(n)$, as happens in the case of Myers-Perry solution to Einstein equations in higher dimensions \cite{Kunduri:2006qa}. This symmetry enhancement allows the sphere sector that corresponds to the spatial slices of the event horizon to be treated as the homogeneous space $S^{2n-1} \simeq SU(n)/SU(n-1)$. Equivalently, $S^{2n-1}$ is the total space of the Hopf fibration, namely an $S^1$ bundle over $\mathbb{CP}^{n-1}$.

The $SU(n-1)$ isotropy representation decomposes as $\mathbf{1}\oplus \mathbf{(n-1)}\oplus \overline{\mathbf{(n-1)}}$. Consequently, the most general $SU(n)$-invariant metric on the base manifold contains only one independent squashing between the Hopf fiber and the $\mathbb{CP}^{n-1}$ base. This enables us to use the stationary metric ansatz
\begin{equation}
\begin{array}{lll}
ds^2 &=&
- b(r)\,\dd t^2 + \displaystyle\frac{\dd r^2}{f(r)} + g(r)\left(\xi^I \xi^I + \widetilde{\xi}^{I} \widetilde{\xi}^{I}\right) \\ [1em]
& & \quad + h(r)\left(\zeta-w(r)\,dt\right)^2 ~,
\end{array}
\label{BHsolution}
\end{equation}
where $I=1,\ldots,n-1$. The $2n-1$ one-forms $\xi^I$, $\widetilde{\xi}^{I}$, and $\zeta$ define a frame adapted to the coset description, constructed from the corresponding Maurer-Cartan form. We denote $\xi^I \xi^I=\xi^I\otimes \xi^I$. The $SU(n)$ symmetry fixes the invariant quadratic combinations
appearing in the metric. The functions $b(r)$ and $f(r)$ determine the temporal and radial components of the metric; namely $b(r)=-g_{tt}+ g^{-1}_{\zeta \zeta }g^2_{\zeta t}$ and $f(r)=g^{-1}_{rr}$, respectively. Function $g(r)=g_{II}$ parametrizes the size of the
$\mathbb{CP}^{n-1}$ base, $h(r)$ describes the squashing of the Hopf fiber, and $w(r)$ is the function associated with the rotation. We subsequently fix the remaining gauge freedom by setting $g(r)=r^2$.

From the viewpoint of physics, the zeroes of $f(r)$ determine the location of the horizons of the black hole solution, with the maximum among its positive roots being the radius of the outer event horizon, $r_H$. The solutions of the equation $b(r)=h(r)w^2(r)$, on the other hand, give the infinite redshift hypersurfaces, which define the limits of ergoregions. $h(r)$ controls how the spatial geometry defined on the planes in which rotation takes place opens up as the radial direction $r$ increases. The function $w(r)$ gives the differential dragging due to rotation. We expect, although we do not impose, that $w(r)$ is an odd function of the angular momentum $a$. This is due to PT invariance under the $\mathbb{Z}_2$ transformations $t\to -t$ and $\zeta \to - \zeta$.

Additional restrictions of the metric functions come from boundary conditions. Consider asymptotically flat boundary conditions for the metric functions, namely
\begin{eqnarray*}
& f(r) \to 1 + \mathcal{O}(r^{-2n+2}) ~, &
~~b(r) \to 1 - \mathcal{O}(r^{-2n+2}) ~, \\ [0.3em]
& w(r) \to \mathcal{O}(r^{-2n}) ~, ~~~~~~~~~&
h(r) \to r^2 + \mathcal{O}(r^{-2n+2}) ~,
\end{eqnarray*}
at large $r$. Then, a black hole exists provided a regular solution satisfying the above boundary conditions has simultaneous zeros of the metric functions $f(r)$ and $b(r)$, namely $f(r_H)=b(r_H)=0$. The angular velocity at the horizon is $w(r_H)=\Omega_H$, while $h(r_H)=\Theta_H$ characterizes the squashing of the sphere at the horizon. To improve the neural network convergence we implement the boundary conditions in terms of a compact radial variable, $r-r_H = \tan(x)$. Further technical details are included in the Appendix.

\section{Results}

\subsection{Rotating black holes in EGB theory}

To test the efficiency of our method, we first solve the $D=5$ case with quadratic Lovelock; \textit{i.e.}, EGB theory. In this case, there is a single parameter controlling the quadratic term; this is $\alpha= 4 \alpha_2/\alpha_1$. These spinning black hole solutions were first constructed in \cite{Brihaye:2010wx} using numerical methods and serve as a test bed for our method. Their approach relies on starting from either the Myers-Perry solution \cite{Myers:1986un} ($\Omega_H \neq  0$, $\alpha = 0$) or the Boulware-Deser solution \cite{Boulware:1985wk} ($\Omega_H=0$, $\alpha \neq 0$) as seeds, and then gradually increase the coupling constant $\alpha$ and/or $\Omega_H\sim \mathcal{O}(a)$ until the desired configuration is reached. In contrast, the PINN training procedure implemented here can construct, from a random initialization, a continuous family of solutions parametrized by $\alpha \in [0,3]$.
\begin{figure}[h]
\centering \includegraphics[width=\linewidth]{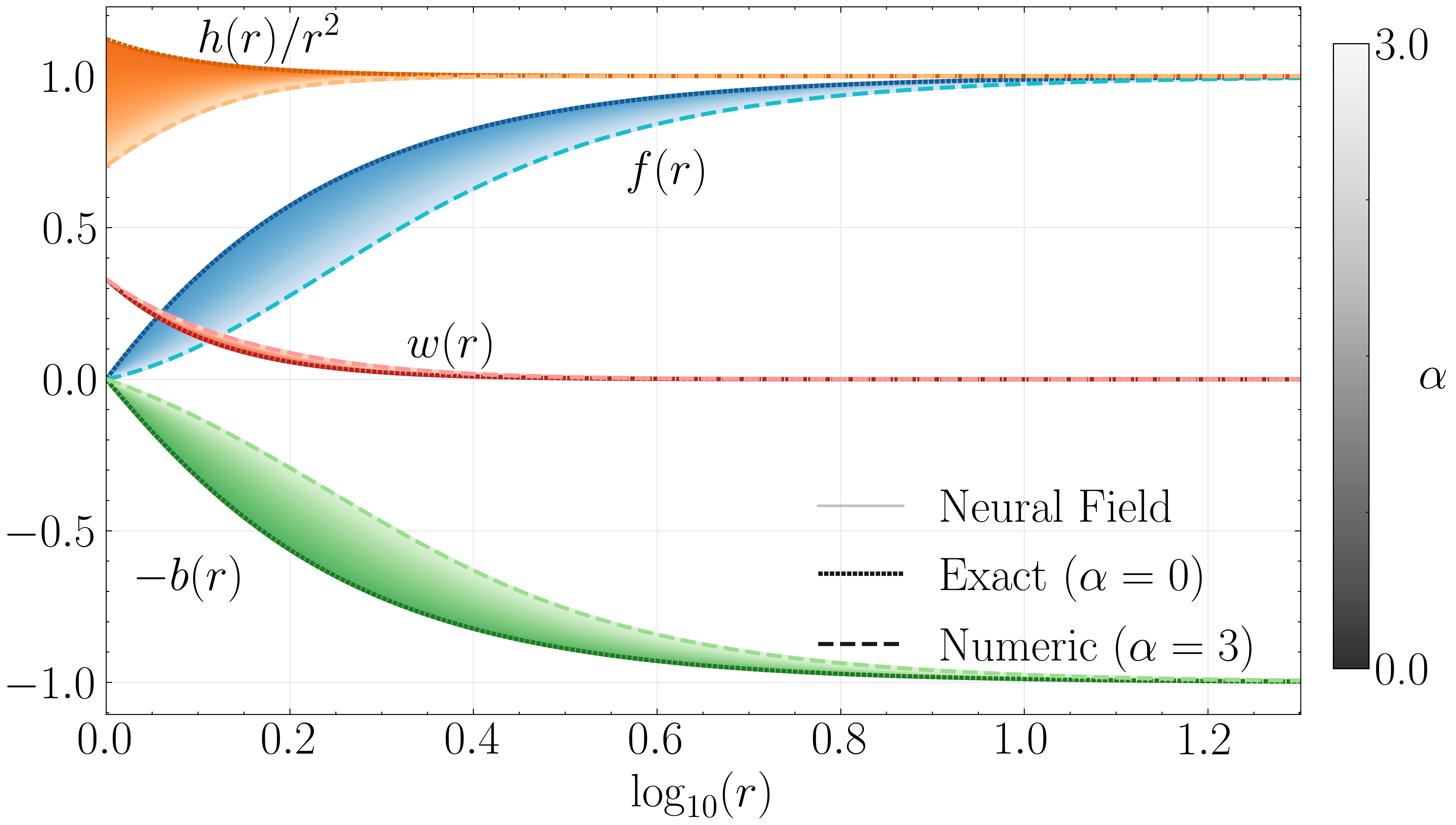}
\caption{Rotating solution with equal angular momenta for $r_H=1, \,\Omega_H = 0.33$ fixed, and $\alpha \in [0, 3]$ in 5D EGB. The gradient bar corresponds to the neural network (NN) value of the parameter input. {\sc Akribeia} neural-spectral solutions reproduce all benchmarks to high precision.}
\label{fig:5D-EGB-solution}
\end{figure}
Figure \ref{fig:5D-EGB-solution} shows a whole family of solutions obtained by our method using a fully connected neural network architecture with three hidden layers of $20$ neurons each. The curves span the coupling range $\alpha \in [0, 3]$ in steps of $0.01$, represented by \textit{solid color-gradient lines}. We immediately notice that for $\alpha=3$ our results agree with \cite{Brihaye:2010wx} (\textit{dashed lines}), while for vanishing $\alpha$ we match the exact Myers-Perry rotating solution (\textit{densely dotted lines}).

There are a few physically significant points to comment regarding the deformations of the metric functions resulting from the inclusion of quadratic terms in conjunction with rotation. First, let us note that the function $h(r)/r^2$ governs the stretching ($h(r)>r^2$) or squashing ($h(r)<r^2$) of the Hopf fiber relative to the $\mathbb{CP}^{n-1}$ base. We observe that $h(r_H)/r_H^2$ decreases with respect to the Myers-Perry case, for the same mass and angular momenta, and there is a critical value of the coupling, $\alpha_\star$, above which the horizon becomes prolate, $h(r_H)/r_H^2 < 1$, suggesting that the solution might be stable against Gregory-Laflamme instabilities typically arising in the ultra-spinning limit \cite{Emparan:2003sy}. Notice that $h(r)$ is monotonic albeit $h(r)/r^2$ is not.

The function $w(r)$ exhibits relatively small deviations from the results of Einstein's theory when compared to the other metric functions. As one might expect, it is monotonic with respect to $r$ for all values of $\alpha$. This is because, as previously mentioned, $w(r)$ accounts for the differential rotation of spacetime as one approaches or moves away from the horizon. This dragging is always co-rotating with the horizon and, for a given value of $\Omega_H$, it increases monotonically with $\alpha > 0$. This suggests that higher-curvature corrections enhance frame dragging throughout the exterior spacetime, which would manifest as a shift in the Lense-Thirring precession rate for orbiting test bodies.

The correlated behavior shown by the functions $f(r)$ and $b(r)$, whose dependence on $\alpha$ for intermediate radii is more apparent, indicates that the effective gravitational radius as well as the gravitational redshift factor grow with respect to their values in Einstein's theory more steeply with $\alpha$. The plot also clearly demonstrates how our neural field parametrizes the solutions in $\alpha$, from the stationary Myers-Perry black hole of Einstein's theory to the numerical stationary solution of EGB \cite{Brihaye:2010wx}.
\begin{figure}[h]
\centering
\includegraphics[width=\linewidth]{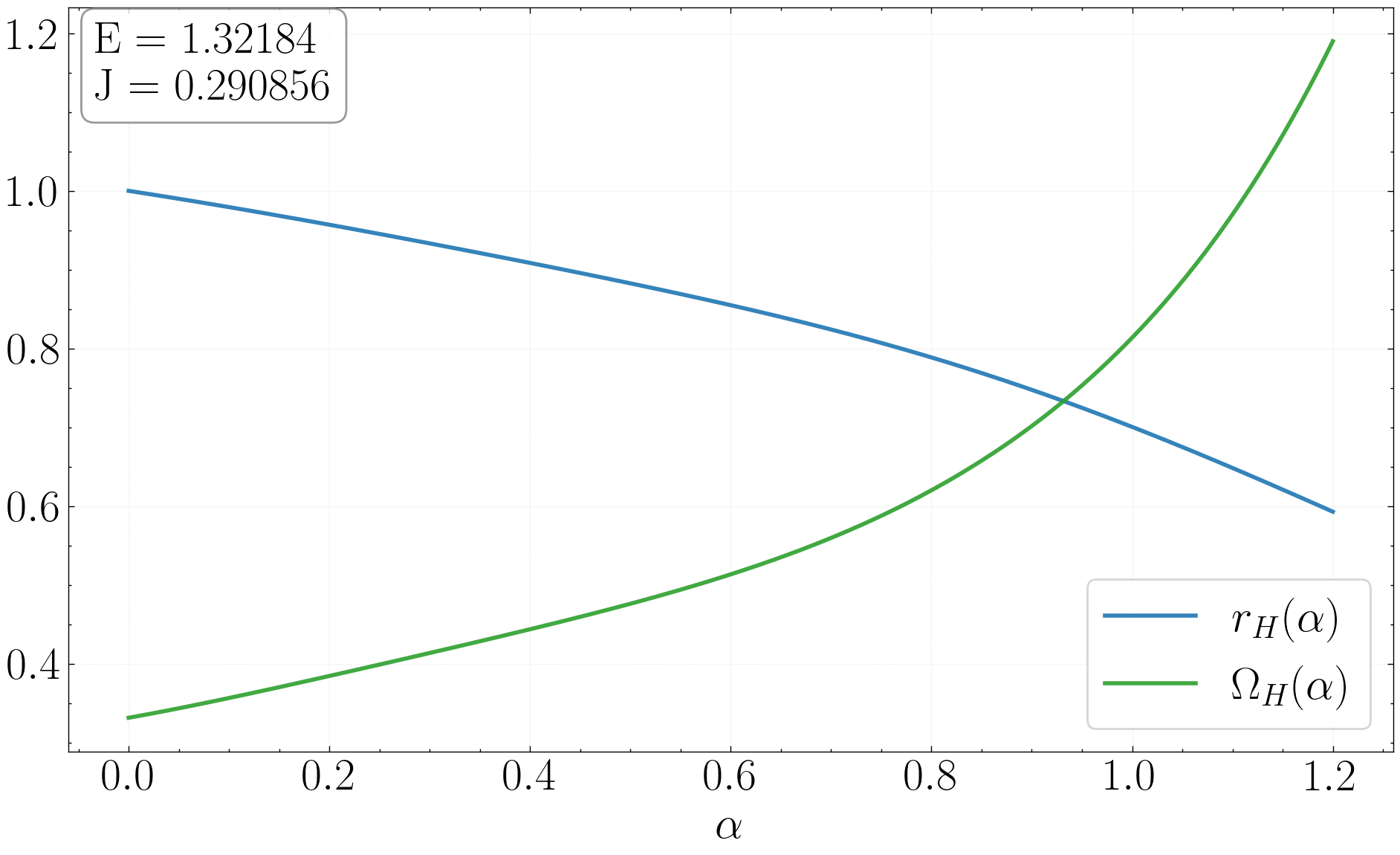}
\caption{Horizon radius $r_H$ and angular velocity $\Omega_H$ as smooth functions of $\alpha$, obtained with {\sc Akribeia} for fixed mass and angular momentum.}
\label{function-of-alpha}
\end{figure}
This serves as a non-trivial consistency check of our hybrid neural-spectral method. A hallmark of {\sc Akribeia} is its ability to trace physical observables continuously across parameter space, as illustrated in Fig. \ref{function-of-alpha}. This continuous parametric control --inaccessible to shooting methods or perturbative approaches-- is precisely what is needed to construct the theoretical templates that gravitational wave and horizon-scale observations will require.

\subsection{Rotating black holes in cubic Lovelock theory}

Having successfully contrasted our method against known results, let us enter into unexplored territory and find for the first time rotating black holes solutions in cubic Lovelock theory. The $D=7$ case, the lowest dimensionality in which those cubic terms actually contribute to the dynamics, a new length scale emerges that competes with the characteristic length $\alpha^{1/2}$. This new scale is $\beta^{1/4}$ with $\beta = 16 \,\alpha_3/\alpha_1$. Even though we fundamentally expect from Effective Field Theory reasoning a single scale, $\alpha^{1/2} \sim \beta^{1/4} \sim \ell$ (the string scale, the Planck length, etcetera), notice that two-scale systems might also lead to interesting phenomena \cite{Thaalba:2025lwe}.

\begin{figure}[h]
\centering
\includegraphics[width=\linewidth]{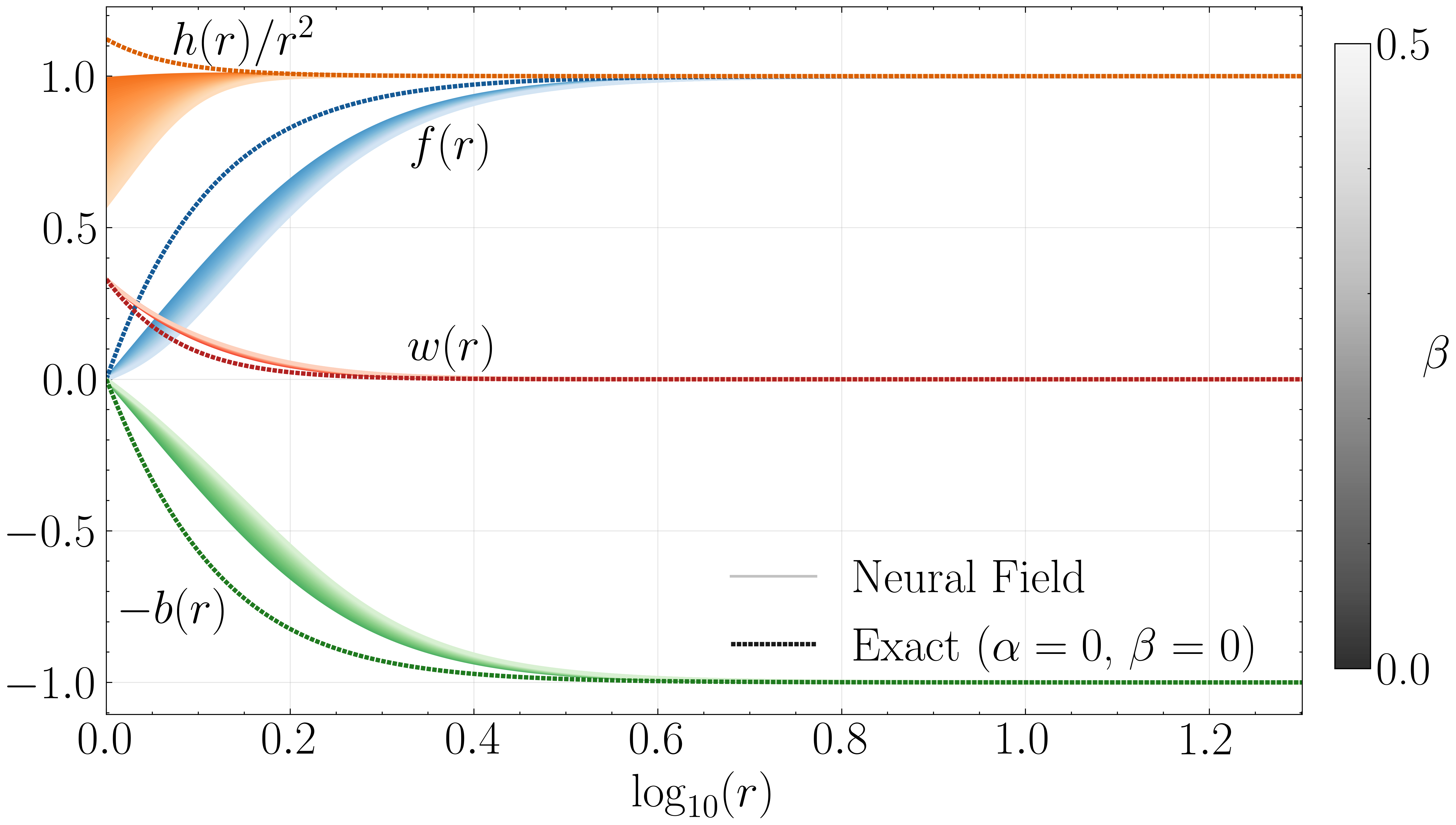}
\caption{Rotating black hole solutions with equal angular momenta for $r_H=1, \,\Omega_H = 0.33$, $\alpha=0.5$ fixed, and $\beta \in [0,0.5]$ in 7D cubic Lovelock gravity.}
\label{fig:7D-Lovelock-solution}
\end{figure}

For the sake of concreteness, the plot in Fig. \ref{fig:7D-Lovelock-solution} has fixed but non-vanishing $\alpha$ and varies only the cubic coupling constant $\beta$. To the best of our knowledge, this is the first reported rotating solution in a theory with cubic curvature terms, featuring finite angular momenta. This is the main result of this paper.

A heuristic consistency check lies in the fact that, when comparing the cubic solution with the quadratic one, one finds qualitatively similar behavior in the metric functions. This demonstrates that we are indeed dealing with the same sector of solutions. The qualitative consequences of adding the cubic term on the stationary sector of black hole solutions is quite similar to the analysis of EGB's case, at least for equal angular momenta. The gaps visible in the graph between the GR solution ($\alpha = \beta = 0$) and the continuous sweep of solutions arise because, as previously mentioned, only $\beta$ is being varied in this plot, while $\alpha$ is fixed and non-vanishing.

Due to the increased difficulty introduced by the cubic term, achieving the same level of accuracy required a deeper network architecture consisting of four hidden layers with $64$ neurons each. The \textit{solid color-gradient lines} here span the cubic coupling $\beta \in [0,0.5]$ in steps of $0.01$, for a fixed value of $\alpha = 0.5$ whereas the \textit{dotted dashed lines} show the analytic Myers-Perry solution in $D=7$ \cite{Myers:1986un}, which our solution consistently approaches in the $\alpha \to 0$ limit. Our hybrid approach successfully passes the initialization sensitivities of spectral solvers, rapidly reducing the residuals far below machine double-precision.

The method generalizes to higher odd dimensions and/or higher-order curvature terms in Lovelock theory. Provided that one considers multiple angular momenta of equal magnitude, the problem of solving the field equations reduces to a system of ordinary differential equations. Although the problem is highly non-linear, 
primarily because second derivatives of the metric tensor appear raised to increasing powers, our hybrid strategy using {\sc Akribeia} makes it possible to find an accurate solution with an ordinary laptop. We will conduct an exhaustive scan of solutions in a forthcoming paper.

\section{Discussion}


We have introduced a hybrid neural-spectral algorithm combining physics-informed neural networks with pseudo-spectral refinement, and applied it to construct rotating black hole solutions in quadratic and cubic Lovelock gravity with equal angular momenta. These are the first fully non-linear rotating solutions in cubic Lovelock gravity for arbitrary values of the gravitational couplings and angular momenta, filling a significant gap in the landscape of exact and numerical solutions in higher-curvature gravity.

The method achieves a striking accuracy, overcoming what is often the main limitation of standard numerical approaches, and all solutions reduce correctly to the known limits as the higher-curvature couplings and/or angular momenta are turned off. The PINN stage provides the high-quality initialization that classical solvers require for convergence, while the pseudo-spectral stage uses extended-precision arithmetic, which is not natively supported by conventional machine-learning libraries.

Several directions open naturally from this work. The most immediate, besides an exhaustive analysis of rotating Lovelock black holes, is a systematic study of linearized perturbations around our solutions to determine whether the suppression of $h(r_H)/r^2_H$ below unity translates into stability against Gregory-Laflamme-type instabilities. On the solution-construction side, {\sc Akribeia} can be extended to higher Lovelock orders, to asymptotically AdS backgrounds ---where the interplay between higher-curvature corrections and superradiant instabilities is particularly rich---, and to other beyond-GR theories admitting a cohomogeneity-1 reduction. It is worth noting that our method may prove useful for solving non-linear differential equations arising in other areas of theoretical physics such as fluid mechanics \cite{GomezSerrano1}.
\section{Future Prospects}

A key step towards astrophysical or cosmological applications demand our method to be adapted to other setups involving a system of coupled non-linear partial differential equations, which we are in the process to complete. That would allow us to make progress in the construction of parametric catalogues of solutions for computing observational signatures ---shadows, photon rings, and gravitational wave ringdown spectra--- that can serve as templates for EHT, LISA, and next-generation detectors.

From an observational standpoint, X-ray probes of accreting black holes offer strong-field tests complementary to horizon-scale imaging and gravitational waves. Black hole spins are routinely measured under the Kerr hypothesis through continuum fitting of the thermal disk spectrum \cite{McClintock:2013vwa}, relativistic reflection spectroscopy \cite{Reynolds:2020jwt} and, most recently, X-ray polarimetry \cite{Krawczynski:2022}. These techniques are chiefly sensitive to the location of the innermost stable circular orbit, to the gravitational redshift, and to frame dragging, i.e., precisely the features that our solutions show to be deformed by the higher-curvature couplings (Figs. \ref{fig:5D-EGB-solution} and \ref{function-of-alpha}).  The enhancement of $w(r)$ translates into shifted Lense-Thirring precession frequencies of the kind invoked to explain quasi-periodic oscillations \cite{Ingram:2019mna}. Certified parametric families of the type delivered by {\sc Akribeia} are exactly what would allow the folding of non-Kerr metrics into these measurement pipelines and to quantify, or even break, the degeneracy between spin and higher-curvature couplings \cite{Bambi:2015kza}.\vskip1mm

\subsection*{Acknowledgments}
We thank Juan \'Alvarez Ruiz and Mat\'\i as Volij for collaboration at early stages of this article, and the authors of \cite{Brihaye:2010wx} for sharing their numerical solution for the sake of comparison. We also thank Julio Oliva, Javier G\'omez Serrano, Tristan Buckmaster, Claudio Mu\~noz and the AstroAI collaboration for interesting suggestions.  J.F.S acknowledges support from NASA Grant NAS8-03060. This work is partially supported by AEI-Spain PID2023-152148NB-I00 and by Maria de Maeztu excellence unit grant CEX2023-001318-M, by Xunta de Galicia (CIGUS Network of Research Centres, projects ED431C-2021/14 and ED431F-2023/19), by the European Union FEDER, by a fellowship from the ”la Caixa” Foundation (ID 100010434) with fellowship code LCF/BQ/PI24/12040029, and by the IGFAE's IGNITE 2025 project sPINNGRAV.\vskip1mm

\subsection*{Author contributions}
\!\!\!\!\! F.A.S., M.O. and A.B. developed the analytical framework and the hybrid numerical scheme, performed the analysis, interpreted the results, drafted part of the manuscript and wrote the Appendix. F.A.S., M.O. and D.A. implemented the open-source code. D.A. designed part of the machine-learning strategy. E.T. assisted in the performance assessment of the machine-learning methods. J.F.S. contributed to contextualizing the results with an observational perspective. J.D.E., G.G. and C.G. conceived the project, interpreted the results, and prepared the final version of the manuscript. All authors discussed the results, reviewed the manuscript, and approved the final version.

\onecolumngrid

\appendix

\section*{APPENDIX}
\subsection{Equations of motion}

The equations of motion of Lovelock gravity in the vielbein formalism can be written as
\begin{equation}\label{field_eqs_lovelock}
    \sum_{k} c_k \epsilon_{a_1 \dots a_{D-1} b} \mathsf{R}^{a_1 a_2} \wedge \dots \wedge \mathsf{R}^{a_{2k-1} a_{2 k}} \wedge V^{a_{2k+1}} \wedge \dots \wedge V^{a_{D-1}} = 0
\end{equation}
where $\mathsf{R}^{a b} = \dd \omega^{a b} + \omega^{a}{}_c \wedge \omega^{c b}$ is the curvature 2-form, $\omega^{a b}$ is the torsionless spin connection, $V^a$ is a vielbein basis, and $c_i$ are the couplings of the different Lovelock terms. For any homogeneous space, the components of the curvature 2-form do not depend on the coordinates when expressed in the vielbein basis. In particular, the horizon geometry of our black holes consists of homogeneous spaces of the form $SU(n)/SU(n-1)$. Under these considerations, the equations of motion in the vielbein basis do not depend on the coordinates of the deformed sphere, rendering the cohomogeneity-1 ansatz consistent.

To define the 1-forms $\xi^{I}, \widetilde{\xi}^{I}$ and $\zeta$ that were used in the explicit construction of the ansatz, let us consider the splitting of the algebra
$\mathfrak{su}(n)=\mathfrak{su}(n-1)\oplus \mathfrak{K}$. We consider the embedding of $\mathfrak{su}(n-1)$ as the lower $(n-1)$ by $(n-1)$ block. A basis of $\mathfrak{K}$ that respects the coset construction is
\begin{align}
    \mathbf{k}_I &= i ( \epsilon_{1, I+1} + \epsilon_{I+1, 1}) \, , \notag \\
    \widetilde{\mathbf{k}}_I &= \epsilon_{1, I+1} - \epsilon_{I+1, 1} \,,
    \qquad I=1,\dots, n-1 \, , \label{basis_K} \\
    \mathbf{p} &= i \sqrt{\frac{2}{n(n-1)}} \left[ (n-1) \epsilon_{1,1}
    - \sum_{I=1}^{n-1} \epsilon_{I+1,I+1} \right] \, , \notag
\end{align}
where $\epsilon_{i,j}$ is an $n$ by $n$ matrix whose only non-vanishing entry is equal to one and is located at row $i$ and column $j$. The matrix $\mathbf{p}$ commutes with $\mathfrak{su}(n-1)$.

The left-invariant Maurer-Cartan 1-form constructed from a coset representative $L \in SU(n)/SU(n-1)$ is decomposed as
\begin{equation}
    \mathbf{\Omega} = L^{-1} \dd L = \xi + \omega \, ,
\end{equation}
where $\omega \in \mathfrak{su}(n-1)$ and $\xi \in \mathfrak{K}$. Their components along the basis \eqref{basis_K} define the 1-forms $\xi^I, \widetilde{\xi}^{I},\zeta$ in the following way:
\begin{equation}
    \xi = \xi^I \mathbf{k}_I + \widetilde{\xi}^I \widetilde{\mathbf{k}}_I + \zeta \, \mathbf{p} \,.
\end{equation}
The Maurer-Cartan equation $\dd \mathbf{\Omega} + \mathbf{\Omega} \wedge \mathbf{\Omega} = 0$ allows us to compute the curvature 2-form without relying on the explicit parametrization of the coset representative $L$.

Hereafter, we impose the gauge condition $g(r)=r^2$. The equation for $w(r)$ can be integrated once, yielding a first-order equation that involves an integration constant $C_0$,
\begin{equation} \label{first_order_eq}
	w'(r) = C_0 \frac{r^{2n-2}\sqrt{b(r)}}{\sqrt{f(r)}h(r)^{3/2} (r^{4n-4} + \mathcal{P}_n(r; f(r), h(r)))}
\end{equation}
where $\mathcal{P}_n$ is a polynomial in $r$, $f(r)$, $h(r)$, and the coupling constants. In the GR limit, one has $\mathcal{P}_n=0$. For quadratic Lovelock in $n=2$ and cubic Lovelock in $n=3$, the polynomials are
\begin{align}
	\mathcal{P}_2 &= -\alpha (r^2 f(r) + 3 h(r) -4 r^2) \, ,\\
	\mathcal{P}_3 &= -6\alpha r^4[r^2 f(r) + h(r) -2 r^2] \\
		&\quad + 9 \beta [r^4 f(r)^2 + 8 r^4 -12 r^2 h(r) 
		 + 5 h(r)^2 + f(r)(2r^2 h(r)-4r^4)] \, . \notag
\end{align}
The full system is too lengthy to display explicitly. It can be found in the following repository \cite{diffEqs}.

The system consists of the first-order equation \eqref{first_order_eq}, one additional first-order equation in $f(r)$, and two second-order equations in $b(r)$ and $h(r)$.

The asymptotic behaviour of the fields as \(r\to\infty\), compatible with the equations~\eqref{field_eqs_lovelock} for the ansatz presented in the main document, is given by
\begin{align}
	f(r) &= 1 + \frac{\mathcal{U}}{r^{2n-2}} + \frac{\mathcal{V}}{r^{2n}} + \frac{n (\mathcal{U V}  + \mathcal{W}^2)  + 2(1-n) \mathcal{U}^2 \frac{c_2}{c_1}}{ 2(1-n) r^{4n-2}} + \mathcal{O}(r^{-4n}) \,, \\
	b(r) &= 1 + \frac{\mathcal{U}}{r^{2n-2}} + \frac{(1-n) \mathcal{U V} + n \mathcal{W}^2 + (2n -1) \mathcal{U}^2 \frac{c_2}{c_1}}{(2n-1)r^{4n-2}} + \mathcal{O}(r^{-4n}) \,, \\
	h(r) &= r^2 + \frac{\mathcal{V}}{r^{2n-2}} - \frac{n}{2(n-1)} \frac{\mathcal{U V} + \mathcal{W}^2}{r^{4n-4}} + \mathcal{O}(r^{-4n+2}) \,, \\
	w(r) &= \frac{\mathcal{W}}{r^{2n}} - \frac{\mathcal{W}(\mathcal{V} - \mathcal{U} \frac{c_2}{c_1})}{r^{4n}} + \mathcal{O}(r^{-6n}) \,.
\end{align}
It is controlled by three real constant $\mathcal{U}$, $\mathcal{V}$, and $\mathcal{W}$ and was verified explicitly for $D=5,7,9,11$.

The boundary conditions at the horizon are
\begin{align}
	f(r) &= f_1 (r-r_H)+ \mathcal{O}(r-r_H)^2 \, , \\
	b(r) &= b_1 (r-r_H)+ \mathcal{O}(r-r_H)^2 \, , \\
	h(r) &= h_H + \mathcal{O}(r-r_H)  \, , \\
	w(r) &= \Omega_H + \mathcal{O}(r-r_H) \, .
\end{align}

Substituting the asymptotic expansion into~\eqref{first_order_eq} yields the relation
\begin{equation}
	C_0 = - 2n \mathcal{W} \, .
\end{equation}

\paragraph{Independent parameters.}
The system is first order in \(f(r)\) and \(w(r)\), and second order in \(b(r)\) and \(h(r)\). It therefore contains six integration constants. The leading-order horizon conditions \(f(r_H)=0\), \(b(r_H)=0\), and \(w(r_H)=\Omega_H\) reduce this number to three, provided the asymptotic condition $w(\infty)=0$ is imposed. In this case, the latter fixes the additional integration constant $C_0$, leaving three independent parameters that match the three free parameters $\mathcal{U}, \mathcal{V}$, and $\mathcal{W}$ appearing in the asymptotic expansion. These may be identified with the horizon data $h(r_H)$, $h'(r_H)$, and $b'(r_H)$.

\noindent At this stage, the counting suggests the existence of a three-parameter family of black-hole solutions. However, this is not the case. As we will demonstrate by using neural networks as a discovery phase, fixing only two parameters (for instance, $\mathcal{U}$ and $\mathcal{W}$, or $r_H$ and $\Omega_H$) is sufficient. The training process naturally determines the third parameter in order to output a regular solution.

\noindent {\bf Analytic preparation of the equations:} The numerical construction of black hole solutions has largely benefited from a previous preparation of the equations. In particular, working with a compact coordinate and making nontrivial boundary conditions leading order has significantly improved the convergence.
We consider the new radial coordinate $x\in [0, \pi/2]$ related to the non-compact radial coordinate $r \in [r_H, \infty)$ in the following manner

\begin{equation} \label{r-compactification}
	r-r_H = \tan(x) \, .
\end{equation}
We consider the following field redefinitions for the metric functions
\begin{align} \label{parametrization-functions}
    f(x) &= 1 + \cos^{2 n - 2}(x) (\sin(x) F(x) - 1) \, , \\
    b(x) &= 1 + \cos^{2 n - 2}(x)(\sin(x) B(x)-1) \,,\\
	h(x) &= (r_H + \tan(x))^2 + \cos^{2n-2}(x)H(x) \, ,\\
	w(x) &= \cos^{2n}(x) W(x) \, ,
\end{align}
The redefinitions of $f$ and $b$ into $F, B$ allow to impose the leading order boundary conditions at the horizon and at infinity in an exact manner. The redefinition of $h$ and $w$ into $H$ and $W$, allow us to imposes the leading boundary condition at infinity.

In this parameterization the relevant constants  of the asymptotic expansion are
\begin{align}
    f_1 &= F(0) \, , \qquad b_1 = B(0) \, , \qquad \Omega_H  = W(0) \, , \qquad h_H = r_H^2 + H(0)  \, \\
    \mathcal{U} &= F(\pi/2) - 1  \, , \qquad \mathcal{W} = W(\pi/2) \, , \qquad \mathcal{V} = H(\pi/2) \, .
\end{align}

Although the parameterization above works in arbitrary odd dimension, in $D=5$, corresponding to $n=2$, an equivalent but simpler choice is available for the metric functions $f$ and $b$. This choice involves a different redefinition of the auxiliary functions $F$ and $B$, but it parametrizes the same class of metric functions. Explicitly, in $D=5$ we use
\begin{align}
f(x) &= \sin (x) \left[1 + \cos^{2} (x) F(x) \right] , \quad
b(x) = \sin (x) \left[1 + \cos^{2} (x) B(x) \right] \, .
\end{align}

In the next section, we introduce the definition of the neural-network ansatz used to solve the equations \eqref{field_eqs_lovelock} for the metric ansatz in dimensions $D=5,7$.

\subsection{Neural Field Approach}
A neural network is a function defined as a composition of affine maps and nonlinear activation functions. Let $\mathcal{V}_i$ be a sequence of vector spaces. For each layer, we define an affine map
\begin{equation}
    \ell_i : \mathcal{V}_i \to \mathcal{V}_{i+1}, \qquad
    \ell_i(h_i)=\mathcal{M}_i h_i + b_i,
\end{equation}
where \(h_i\in\mathcal{V}_i\), \(\mathcal{M}_i\) is a constant matrix, and \(b_i\) is a constant vector. The \textit{activation function} \(\sigma_i\) acts component-wise on \(\mathcal{V}_{i+1}\). The neural network is then given by the composition
\begin{equation}
    \mathcal{N}: \mathcal{V}_1 \xrightarrow{\sigma_1 \circ \ell_1} \mathcal{V}_2 \xrightarrow{\sigma_2 \circ \ell_2} \dots \longrightarrow \mathcal{V}_{L-1} \xrightarrow{\sigma_L\circ \ell_L}\mathcal{V}_L \, ,
\end{equation}
Equivalently, one may write $h_{i+1}=\sigma_i\!\left(\ell_i(h_i)\right)$ for $i=1,\dots,L-1$. The collection of matrices and bias vectors defines the trainable parameters of the network,
\begin{equation}
    \mathbb{W}=\{(\mathcal{M}_i,b_i): i=1, \dots , L-1\} \, .
\end{equation}
In computer science, this architecture is usually referred to as a multilayer perceptron (MLP) or a fully connected neural network (FCNN). Denoting $n_i=\dim(\mathcal{V}_i)$, we represent the architecture by the list $[n_1,n_2,\dots,n_L]$.

Provided the activation functions are differentiable, the network is differentiable with respect to both its inputs and its parameters. This makes it suitable for approximating physical quantities governed by differential equations and boundary conditions. In the framework of physics-informed neural networks, the solution is obtained by minimizing an object called the \textit{loss function}, which takes the form
\begin{equation}
    \mathcal{L}(\mathbb{W}) = \omega_{1}\mathcal{L}_{\mathrm{equations}}(\mathbb{W}) +\omega_{2}\mathcal{L}_{BC}(\mathbb{W}),
\end{equation}
where $\omega_1, \omega_2$ are positive parameters allowing to balance each term  of the loss. For differential equations, the main term in the loss function is the mean square residual 
\begin{align}
    \mathcal{L}_{\mathrm{equations}}(\mathbb{W}) &= \frac{1}{N}\sum_{\mathbf{x} \in \mathcal{I}} |\mathcal{E}\left[\mathcal{N}(\mathbf{x}),\nabla\mathcal{N}(\mathbf{x}),...\right]|^2,
\end{align}
where the set $\mathcal{I} \subset \mathcal{V}_1$ is a discrete and finite set of points of the input layer called \textit{collocation points} and $N$ is the cardinality of the set $\mathcal{I}$. {$\mathcal{E}$ are the differential equations that depend on the neural network and its derivatives with respect to the points. In our problem, the domain is given by an appropriate subset of the plane $\mathbf{x}=(x,\alpha)\in \mathbb{R}^2$.

A second contribution to the loss function is often included to impose prescribed values of the solution on a subset \(\Sigma\subset\mathcal{V}_1\), typically a lower-dimensional set such as a boundary. We define

\begin{align} \label{loss-soft-condition}
\mathcal{L}_{BC}(\mathbb{W}) &= \frac{1}{\widetilde{N}}\sum_{\mathbf{x}_0 \in \widetilde{\mathcal{I}}} |\mathcal{N}(\mathbf{x}_0)-\mathcal{F}(\mathbf{x}_0)|^2
\end{align}
where $\widetilde{\mathcal{I}}$ is a discrete subset of $\Sigma$, $\widetilde{N}$ its cardinality and $\mathbf{x}_0 = (x_0, \alpha) $. The subscript $BC$ stands for boundary conditions.

When boundary conditions are imposed along a single coordinate axis (e.g., $x = x_0$), the output of the neural network can be analytically reparametrized to satisfy the condition \(\mathcal{N}(\mathbf{x}_0)=\mathcal{F}(\mathbf{x}_0)\) exactly. To achieve this, we redefine the network output as:
\begin{equation}
   \widehat{\mathcal{N}}(\mathbf{x}; \mathbb{W})=\mathcal{F}(\mathbf{x}_0)+\left(1-e^{-(x-x_0)}\right)\mathcal{N}(\mathbf{x};\mathbb{W})\,.
\end{equation}
By construction, \(\widehat{\mathcal{N}}(\mathbf{x}_0)=\mathcal{F}(\mathbf{x}_0 )\). We refer to this implementation of the boundary condition as \textit{strong}.

\subsection{{\sc {\bf Akribeia}}: a discovery-refinement scheme}
Here, we provide details regarding the hybrid scheme, {\sc Akribeia} \cite{akribeia}, presented in the main text. This is divided into two primary phases:
 
 \begin{itemize}
     \item {\bf Discovery phase:} A PINN is used to find an approximate solution to the boundary value problem starting from random initialization.

    \vspace{5mm}
     
     \item {\bf Refinement phase:} A pseudo-spectral method is initiated with the output of the PINN, to provide a high-precision numerical solution.
 \end{itemize}


\subsubsection{Discovery Phase: Neural Network}

This phase comprises a series of stages that are used to find the initial seed for the refinement phase.

\noindent {\bf Training strategy:} The primary computational framework is implemented in \textit{Pytorch}, utilizing the \textit{Pytorch lightning} interface. The code \cite{difflearning} is designed to solve various systems of differential equations using PINNs. The framework is modular, strictly separating the definition of the physical problem from the standard training/optimization procedure. The general training configurations employ FCNNs architectures and a single activation function across all hidden layers.

\noindent {\bf Random initialization:}  Given the architecture of the neural network, the remaining free parameters are the entries of the weight matrices $\mathcal{M}_i$ and the bias vectors $b_i$. We use three different initializations:
\begin{itemize}
    \item {\sc Xavier uniform}: the weight matrices are initialized with Xavier uniform
    initialization \cite{Xavier,PytorchDocumentation}. The biases are sampled from a uniform distribution centered at zero with radius $1/\sqrt{n_i}$.
    
    \item {\sc Kaiming-like}: the weight matrices of all hidden layers, excluding the final layer, are initialized with Kaiming uniform initialization \cite{Kaiming,PytorchDocumentation}, using \nolinkurl{mode="fan\_in"} and \nolinkurl{nonlinearity="relu"}.  The final layer is initialized separately and sampled from a uniform distribution centered at zero with radius $0.05/\sqrt{n_{L-1}}$. All bias vectors are initialized to zero. This choice produces a small initial output of the network.
    \item {\sc Orthogonal}: the weight matrices are initialized as
    semi-orthogonal matrices \cite{Orthogonal,PytorchDocumentation}. The biases are initialized as zero vectors.
\end{itemize}

We initialize the models randomly using one of the initialization mentioned above.

\noindent {\bf Sampling points:}  To sample points in a domain of dimension greater than one, for instance an open subset of $\mathbb{R}^n$ with $n>1$, the Cartesian-product strategy becomes increasingly expensive, since the total number of points grows as $N^n$, where $N$ is the number of points along each direction. Instead, for each variable we generate a one-dimensional array of $N$ points according to its prescribed distribution. These arrays are then independently shuffled using pseudo-random permutations. The $n$-dimensional sample is constructed by pairing the first elements of all arrays, then the second elements, and so on. In this way, the domain is populated with $N$ points rather than $N^n$, while preserving the  prescribed one-dimensional distributions for each variable.

In our case, the inputs are the coordinate $x$ and the corresponding coupling constant $\alpha$. To efficiently navigate the highly non-linear loss landscape, stabilize the training, and better resolve boundary gradients, we divide the training protocol into \textit{two main optimization stages}, typically requiring $10^4$ to $10^5$ steps. The first stage uses $N_1$ collocation points and relies on stochastic, first-order momentum-based methods. The second stage performs a local deterministic refinement using a quasi-Newton second-order method, increasing the number of collocation points to $N_2$, with $N_2 \gtrsim 2N_1$. This progressive sampling strategy improves the accuracy of the resulting solution. The complete algorithm is structured as follows:

\noindent {\bf Stage 1: Global Exploration:} In the initial stage, the network weights and biases are optimized using \textit{Adam}. We selected this first-order gradient-based stochastic optimizer for its robustness against noise gradients and its ability to avoid early entrapment in local minima. The learning rate during this stage is governed by a multi-stage scheduler: 
\begin{itemize}
    \item {\bf Warmup:} The learning rate increases linearly from $10^{-6}$ to a maximum of $10^{-3}$. This prevents large and destabilizing gradient updates during initial iterations when random initialization yields high residuals, facilitating rapid convergence to an $\mathcal{O}(1)$ loss.
        
    \item {\bf Stable:} The learning rate remains constant at $10^{-3}$, allowing the network to explore the loss landscape for a sufficiently large number of steps and to identify a favorable valley near the target solution.
    
    \item {\bf Decay:} Once the loss stagnates, we use PyTorch's \textit{ReduceLROnPlateau} scheduler to lower the learning rate. This reduces stochastic fluctuations and helps the network settle deeper into the identified minimum. After a satisfactory reduction during this plateau phase, we apply a faster, progressive learning-rate decay using a cosine annealing scheduler.
\end{itemize}

\vspace{3mm}

\noindent {\bf Stage 2: Local Optimization (L-BFGS):} Once Adam reaches a saturation point in minimizing the physical residuals, the training transitions to the \textit{L-BFGS} optimizer. This is a quasi-Newton method leverages second-order information (Hessian approximations) to accelerate convergence, achieving loss values of $\mathcal{O}(10^{-5})$.



\subsubsection{Refinement Phase: Pseudo-Spectral Method}

	This phase of the hybrid scheme combines a pseudo-spectral discretization~\cite{Boyd} with a Gauss-Newton optimizer \cite{Nocedal}. The pseudo-spectral method transforms the system of ODEs into an algebraic system, while the Gauss-Newton step iteratively approaches the solution of that system. Once a high-precision solution is attained, this is written in the series of Chebyshev polynomials using a Discrete Chebyshev Transform \cite{Canuto}, while keeping high precision. See \cite{Biasi, Maliborski} for the effectiveness of such methods in Euler equations and in an Einstein-Klein-Gordon model.
	
	\noindent {\bf Pseudo-spectral discretization:} The discretization is performed on a set of collocation points $\{x_n\}$ derived from Chebyshev points of the first kind via $x_n = \frac{\pi}{4}(z_n + 1)$, where
	\begin{equation}
		z_n = \cos\!\left(\frac{\pi}{2N}(2n-1)\right), \qquad n = 1, \ldots, N,
	\end{equation}
	and $N$ is the total number of collocation points. The values of the fields at those points are denoted following the rule $F_n = F(x_n)$, and derivatives in $x$ are approximated by the differentiation matrix
	\begin{equation}
		D_{ij} =
		\begin{cases}
			\dfrac{2 x_j}{\pi(1 - x_j^2)} & i = j, \\[6pt]
			(-1)^{i+j}\dfrac{4\sqrt{(1-x_j^2)/(1-x_i^2)}}{\pi(x_i - x_j)} & i \neq j.
		\end{cases}
	\end{equation}
    With this discretization, the problem can be formulated in two ways, depending on which quantity is held fixed during the optimization process.
    \begin{itemize}
    \item The first possibility consists of fixing $C_0$ and solving the system for $\{F(x),B(x),H(x)\}$. The equation for $w(x)$ decouples from the others after substituting $w'$ with (\ref{first_order_eq}) in the rest of the equations. After discretization, the system of ODEs reduces to an algebraic system of $3N$ equations for the $3N$ unknowns $\{F_n,B_n,H_n\}$ that can be solved for fixed values of $C_0$, independently of $w(x)$. Such metric function is subsequently constructed by directly solving the linear system of equations resulting from the discretization of (\ref{first_order_eq}). The boundary condition $w(x=\pi/2)=0$ is imposed through the barycentric Lagrange interpolation \cite{Barycentric}
	\begin{equation} \label{bary_lagrangian_inter}
		w_{\mathrm{int}}(x) =
		\frac{\displaystyle\sum_{n=1}^{N} \frac{\tilde{p}_n}{x - x_n} w_n}
		{\displaystyle\sum_{n=1}^{N} \frac{\tilde{p}_n}{x - x_n}},
		\qquad \tilde{p}_n = (-1)^{n-1} \sin\!\left(\frac{\pi}{2N}(2n-1)\right),
	\end{equation}
    by replacing one equation of the linear system with $w_{\mathrm{int}}(\pi/2)=0$, so that the number of independent equations matches the number of unknowns.

    \item The other possibility consists of fixing the value of the function $W(x=0)=\Omega_H$ at the horizon and including $C_0$ as an additional unknown to be determined. In this case, the system of four ODEs becomes an algebraic system of $4N$ equations for the $4N+1$ unknowns $\{F_n,B_n,H_n,W_n,C_0\}$. The additional boundary-condition equation is obtained by using the barycentric Lagrange interpolation in equation (\ref{bary_lagrangian_inter}) for $W_n$ to evaluate the function at the horizon. It is then added to the system, resulting in $4N+1$ equations for $4N+1$ unknowns.

    \end{itemize}

    \noindent {\bf Gauss-Newton optimizer:}  Let us denote by $R_i$ the equation residuals evaluated at the collocation points. The Gauss-Newton update of the variables, which are denoted in the form of an array $v_i$ reads 
	\begin{equation}
		v_i^{(\text{new})} = v_i^{(\text{current})} - \lambda\, J_{ij}^{-1} R_j,
		\qquad J_{ij} = \frac{\partial R_i}{\partial v_j},
	\end{equation}
    where $\lambda$ is a step-size parameter to be adjusted on $(0,1]$. For instance, for the formulation with $C_0$ fixed $v_i = (\{F_n\}, \{B_n\}, \{H_n\})_i$ and the array of equations evaluated at the collocation points is $R_i= ((\text{Eq}_1)_n,(\text{Eq}_2)_n,(\text{Eq}_3)_n)_i$. When the formulation of $4N+1$ equations with $\Omega_H$ fixed is used, the Jacobian must be adapted accordingly.

    \noindent {\bf Refinement procedure:} The iterative solver is initialized using the approximate solution generated by the PINN on a moderate collocation grid with $N = 20$. A refinement procedure is then implemented by progressively increasing the number of collocation points in the sequence $N=50,\ 100,\ 200$, using the solution at each resolution as the initial seed for the next one. The method is
	implemented using extended-precision arithmetic to achieve the high precision reported in this work. The use of such a library does not drag the performance of the method, obtaining residuals of order $10^{-80}$ for $N=200$ and 300-digit numbers in runtimes of about one minute. 

    \noindent {\bf Discrete Chebyshev Transform:} This refinement stage concludes by writing the approximate solution on the basis of Chebyshev polynomials of the first kind:
    \begin{equation}
        F_{\text{numeric}}(x) = \sum_{n=0}^{N-1} c_n\, T_n\left(\frac{4}{\pi}x-1\right), \qquad \text{with} \qquad T_n(z) = \cos(n \arccos(z)), \quad z\in [-1,1],
    \end{equation}
   and
    \begin{equation}
        c_n = \frac{p_n}{N} \sum_{k=1}^{N}  (-1)^n \cos\left((2k-1) n \frac{\pi}{2N}\right)\,F_k, \qquad \text{with} \quad p_n = 2, \quad \text{except for} \quad p_{0}=1.
    \end{equation}
    These differentiable functions allow exact evaluation of the equations at any point. 

\subsection{Tests and Results: high-precision from random initialization}

As discussed above, specifying two parameters is sufficient to find a regular black hole solution. In our main approach to solving the system of equations \eqref{field_eqs_lovelock}, we choose to fix the horizon radius at $r_H = 1$ and the horizon angular velocity at a value we denote by $\Omega_H$. The latter is enforced using the strong boundary condition approach to set $W(0) = \Omega_H$. By doing so, the framework converges to a solution. We use this method for the majority of the results presented below, and discuss an alternative approach in the last subsection.

\subsubsection{Test Myers-Perry solution}

The discovery-refinement protocol has been first tested on the Myers-Perry solution ($\alpha=0$), which admits an explicit expression for direct comparison:
\begin{align}
	f(r) &= 1 - \frac{1}{1-\rhOh}\rhoverr^{2n-2} + \frac{\rhOh}{1-\rhOh}\rhoverr^{2n} \, ,\\
	b(r) &= 1 - \rhoverr^{2n-2}\frac{1}{1-(1-\rhoverr^{2n})\rhOh} \, ,\\
	h(r) &= r^2\left[ 1 + \rhoverr^{2n} \frac{\rhOh}{1-\rhOh} \right] \, ,\\
	w(r) &= \rhoverr^{2n} \frac{\Omega_H}{1-\left[1-\rhoverr^{2n}\right]\rhOh} \, .
\end{align}
Starting from different random initializations, our protocol provides approximate solutions with relative errors much below machine-double precision. The precision of the method is quantified by the combination of the relative errors to the analytic expression for each metric function:
\begin{equation}
    \Delta_{\text{Total}} = \Delta_f + \Delta_b + \Delta_{\tilde{h}} + \Delta_\omega, \qquad \text{with} \qquad \Delta_g := \frac{\|g_{\text{numeric}}(x)-g_{\text{analytic}}(x)\|_{L^2}}{\|g_{\text{analytic}}(x)\|_{L^{2}}},
\end{equation}
where the $L^2$-norm is computed in out-of-collocation points of $M\sim 10^{3}$  as follows:
\begin{equation}\label{L2-norm}
    \|g(x)\|_{L^2} = \left(\int_{0}^{\pi/2} g(x)^2 dx\right)^{1/2} \approx \left(\frac{\pi}{2 M} \sum_{m=0}^{M} g(x_m)^2\right)^{1/2}  \qquad \text{with} \qquad x_m = \frac{\pi}{2 M}m.
\end{equation}
This test has been conducted for the cases $n=2$ and $n=3$ for the random initialization defined above: Xavier uniform, Orthogonal, and Kaiming-like, observing no relevant difference between the results; see Table~\ref{tab:Myers-Perry_Solutions}. The comparison of the different phases of the {\sc Akribeia} protocol with respect to the analytical solution of Myers-Perry with $n=2$ is shown in Figure~\ref{fig:Myers-Perry_solution}.


\begin{table}[h]
\centering
\setlength{\tabcolsep}{8pt}
\renewcommand{\arraystretch}{1.2}

\begin{tabular}{|c||c|c|c|c|c|c|c|}
\hline
\multicolumn{8}{|c|}{Xavier uniform random initialization} \\ \hline
Dimension & $\Omega_H$ & 0.1 & 0.2 & 0.3 & 0.4 & 0.5 & 0.6 \\ \hline
 $5$ & $\mathcal{O}(\Delta_{\text{Total}})$ &
$10^{-104}$ & $10^{-100}$ & $10^{-98}$ &
$10^{-96}$ & $10^{-94}$ & $10^{-92}$ \\ \hline
$7$ & $\mathcal{O}(\Delta_{\text{Total}})$ &
$10^{-96}$ & $10^{-92}$ & $10^{-89}$ &
$10^{-87}$ & $10^{-85}$ & $10^{-83}$ \\ \hline
\end{tabular}

\vspace{0.3cm}

\begin{tabular}{|c||c|c|c|c|c|c|c|}
\hline
\multicolumn{8}{|c|}{Orthogonal random initialization} \\ \hline
Dimension & $\Omega_H$ & 0.1 & 0.2 & 0.3 & 0.4 & 0.5 & 0.6 \\ \hline
$5$ & $\mathcal{O}(\Delta_{\text{Total}})$ &
$10^{-104}$ & $10^{-100}$ & $10^{-98}$ &
$10^{-96}$ & $10^{-94}$ & $10^{-92}$ \\ \hline
$7$ & $\mathcal{O}(\Delta_{\text{Total}})$ &
$10^{-96}$ & $10^{-92}$ & $10^{-89}$ &
$10^{-87}$ & $10^{-85}$ & $10^{-83}$ \\ \hline
\end{tabular}

\vspace{0.3cm}

\begin{tabular}{|c||c|c|c|c|c|c|c|}
\hline
\multicolumn{8}{|c|}{Kaiming-like uniform random initialization} \\ \hline
Dimension & $\Omega_H$ & 0.1 & 0.2 & 0.3 & 0.4 & 0.5 & 0.6 \\ \hline
$5$ & $\mathcal{O}(\Delta_{\text{Total}})$ &
$10^{-104}$ & $10^{-100}$ & $10^{-97}$ &
$10^{-96}$ & $10^{-94}$ & $10^{-92}$ \\ \hline
$7$ & $\mathcal{O}(\Delta_{\text{Total}})$ &
$10^{-96}$ & $10^{-92}$ & $10^{-89}$ &
$10^{-87}$ & $10^{-85}$ & $10^{-84}$ \\ \hline
\end{tabular}

\caption{Relative error to Myers-Perry solutions of $r_H=1$ for three distinct random initializations.}
\label{tab:Myers-Perry_Solutions}
\end{table}


\subsubsection{Quadratic Lovelock}
To further demonstrate the accuracy of our method, we present the results reproduced from \cite{Brihaye:2010wx} for the 5D rotating black hole. Table \ref{tab:EGB5D solution} summarizes the specific configuration used to obtain the solution presented in the main text, while Figure \ref{fig:loss_EGB_5D_Oh_0.33} shows the corresponding solution and training loss.

\noindent In addition to the findings presented in the main text, our framework accurately reproduces the solutions reported in \cite{Brihaye:2010wx} for $\alpha =2$, $\Omega_{H} = 0.68$. The results and the respective loss function are displayed in Figure \ref{fig:EGB_5D_Oh_0.68} for $\Omega_H = 0.68$ and $\alpha \in [0, 2.0]$. A feature of the solution noticed here is that, for a given value of $(r, f(r))$, there could be more than one solution that crosses that point for some value of $\alpha$. In other words, for increasing values of $\alpha$ for $\Omega_H=0.68$, the functions $f(r)$ and $b(r)$ initially increase relative to the Myers-Perry solution up to a critical value of $\alpha$. Beyond this point, the functions decrease until they match the target numerical data. This behaviour contrasts with what is observed in the solution for $\Omega_H =  0.33$.

We display in Figure \ref{fig:EGB_phases_solution} the lower-case functions, the upper-cases functions and the residuals in different phases of the {\sc Akribeia} protocol. 


\subsubsection{Cubic Lovelock}

The main result of the article corresponds to the construction of rotating solutions in Lovelock theory in $D=7$. One novelty of this theory is that it has two dimensionful coupling constants $\alpha$ and $\beta$. For the sake of concreteness we have fixed $\alpha$ and considered $\beta$ as one of the inputs parameters of the neural network. 

\noindent The seven dimensional system is more involved than the five dimensional one. This can be corroborated by looking at the length of the equations. Due to this fact, in order to achieve a loss function of order $\mathcal{O}(10^{-4}) - \mathcal{O}(10^{-5})$, we had to increase the depth of the neural network to [2, 64, 64, 64, 64, 1] with respect to the 5D case. This increases the computational cost and computational time. We detail this in the next section.

\noindent Table \ref{tab:7Dsolution} summarizes the configuration for the solution discussed in the main text. The corresponding metric and loss functions are shown in Figure \ref{fig:loss_Lovelock_7D}, while Figure \ref{fig:lovelock_7D_Oh_0.6} presents an additional case for $\Omega_H = 0.6$.

\noindent To demonstrate the accuracy of {\sc Akribeia}, Figure \ref{fig:Lovelock_Phases_solution} displays the upper-case and lower-case metric functions alongside the equation residuals at each phase of the protocol. Focusing on the bottom row, the pseudo-spectral refinement phase reduces the total residual to $\mathcal{O}(10^{-80})$. We achieved this precision by extending the Gauss-Newton continuation schedule, summarized up to $N=200$ in Table \ref{tab:7Dsolution}, in increments of $\Delta N = 50$ until $N=500$. As illustrated in Figure \ref{fig:residual_refinement}, the method exhibits exponential convergence to the solution as we increase the number of collocation points in the refinement phase. We quantify the performance by considering the $L^2$-norm for the residuals defined in Eq. \eqref{L2-norm} where $g(x) = \mathcal{R}(x)$, the corresponding differential equation evaluated in $x$. We compare the total residual of the differential equations defined as
\begin{equation}
    \text{Res}_{\text{Total}} = \text{Res}_{\text{eq}_1} + \text{Res}_{\text{eq}_2} + \text{Res}_{\text{eq}_3} + \text{Res}_{\text{eq}_4},\qquad \text{Res} := \Vert{}\mathcal{R}(x)\Vert{}_{L^2} \, , \label{total_L2_residual}
\end{equation}
for the Lovelock solution constructed with different number of collocation points.

\begin{figure}
    \centering
    \includegraphics[width=0.4\linewidth]{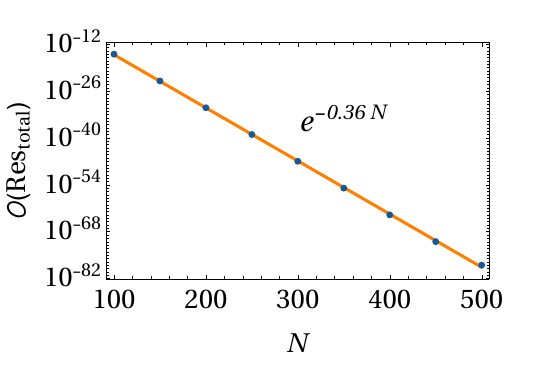}
    \caption{Order of magnitude of the total residual of the 7D equations, as defined in \eqref{total_L2_residual}, for the model with $\alpha=0.5$ and $\beta=0.4$, as a function of the number of collocation points used in the refinement stage. The continuous line corresponds to the fit of a exponential decay with the number of collocation points.}
    \label{fig:residual_refinement}
\end{figure}

\subsubsection{Alternative approach}

In order to study the effect of higher-order corrections on the horizon quantities $r_H$ and $\Omega_H$, an alternative approach can be taken to solve the system of equations. This approach consists of fixing the asymptotic parameters $\mathcal{U}$ and $\mathcal{W}$, which strictly determine the conserved spacetime energy $E$ and angular momentum $J$. Specifically, in five dimensions ($D=5$), these charges are given by
\begin{equation}
E = -\frac{3V_3}{16\pi G} \mathcal{U} \,, \qquad J = \frac{V_3}{8\pi G} \mathcal{W} \,,
\end{equation}

where $V_3 = 2\pi^2$ is the area of the unit 3-sphere. This is the framework used to generate the results shown in Figure 2 of the main text. To select the target values for $E$ and $J$, we use the known analytical relations for the Myers-Perry solution, whose asymptotic parameters are
\begin{equation}
\mathcal{U} = - \frac{r_H^{2n -2 }}{1 - r_H^2 \Omega_H^2} \,, \qquad \mathcal{W} = \frac{\Omega_H r_H^{2n}}{1 - r_H^2 \Omega_H^2} \,, \qquad \mathcal{V} = \frac{\Omega_H^2 r_H^{2n + 2}}{1 - r_H^2 \Omega^2_H} \,.
\end{equation}
By evaluating these expressions for $n=2$ with a horizon radius $r_H=1$ and an angular velocity $\Omega_H=0.33$, we fix the conserved charges of the system to $E= 1.32184$ and $J=0.290856$. Under these conditions, we observe that the effect of increasing the coupling constant is to decrease the horizon radius while simultaneously increasing the angular velocity of the black hole, until the extremal limit ($T=0$) is reached.

\subsection{Computational costs}

The proposed discovery-refinement method was implemented on commercial laptops, with no use of high-performance computing resources. The discovery phase has been implemented in two different machines: one using  NVIDIA RTX 5070 GPU acceleration and the other one using pure CPU computing with a INTEL CORE ULTRA 7. In the 5-dimensional case the performance is similar for both of them, however, in the 7-dimensional case, in particular in the L-BFGS stage, the GPU performs about two times faster than the pure CPU. As a reference of the computational costs, neural-field construction for 5D-EGB black holes with $\alpha \in [0,3]$, using FCNNs of architecture $[2, 20,20,20, 1]$ and $6000$ collocation points, requires on average $15$--$120$ minutes of wall-clock time, depending on the desired level of refinement, with training typically performed for $10^{4}$--$10^{5}$ epochs. The subsequent refinement phase, using the pseudo-spectral method, employs $N=200$ collocation points and converges beyond machine precision in approximately one minute of wall-clock time per solution. This is implemented using a single-core CPU. 

\subsection{Open-source implementation}


To facilitate reproducibility and further development, we provide an open-source implementation of the {\sc Akribeia} framework \cite{akribeia}. It is based on two main open-source repositories developed in this work: the PINN-based discovery stage \cite{difflearning}, and  the pseudospectral refinement procedure \cite{gauss-newton-refiner}. We hope that releasing these tools will provide a useful starting point for future investigations employing hybrid machine-learning and spectral approaches to the solution of boundary-value problems in gravity and other fields of physics and mathematics.


\begin{figure}[h!]
    \centering
    \includegraphics[width=1.0\linewidth]{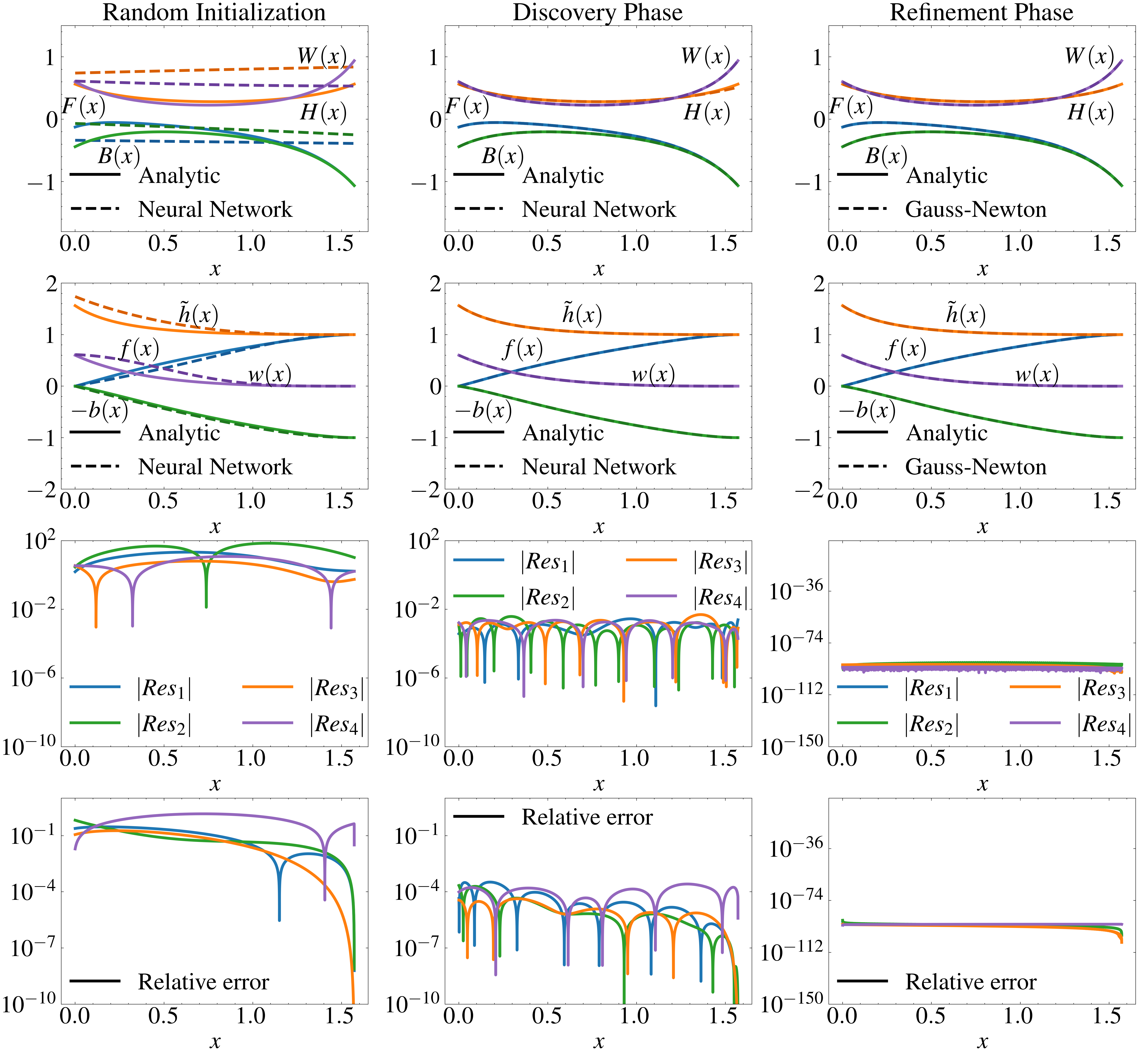}
    \caption{Row 1: upper-case functions of $x$. Row 2: lower-case functions of $x$. Row 3: equation residuals. Row 4: relative errors between the analytical solution and the lower-case functions obtained at each phase. Each column compares one of the three phases of the method for the 5D Myers-Perry solution with $\Omega_H = 0.6$. Phase 0 shows the network output after random initialization (Random Initialization), Phase 1 the output after the neural-network \textit{L-BFGS} optimization (Discovery Phase), and Phase 2 the output after the Spectral Solver optimization (Refinement Phase).}
    \label{fig:Myers-Perry_solution}
\end{figure}


\begin{figure}[h!]
    \centering
    \includegraphics[width=1.0\linewidth]{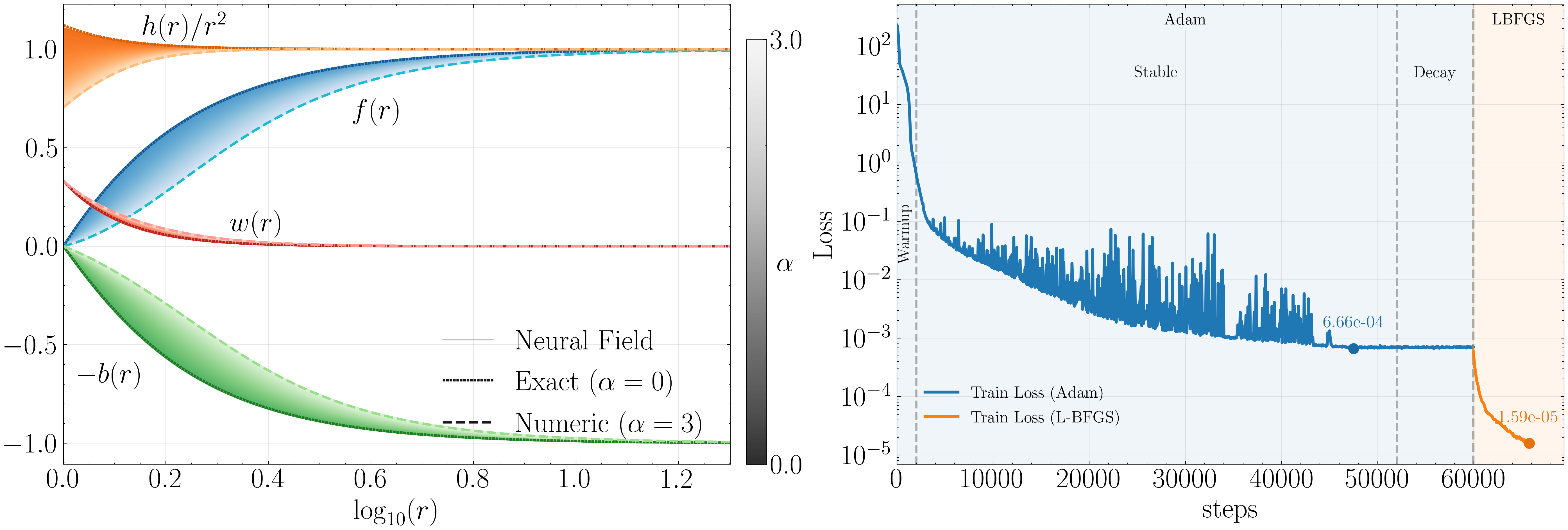}
    \caption{EGB 5D parametric solution for $\Omega_H = 0.33$ with $\alpha \in [0, 3.0]$ presented in the main text and the corresponding loss function. }
    \label{fig:loss_EGB_5D_Oh_0.33}
\end{figure}

\begin{figure}[h!]
    \centering
    \includegraphics[width=1.0\linewidth]{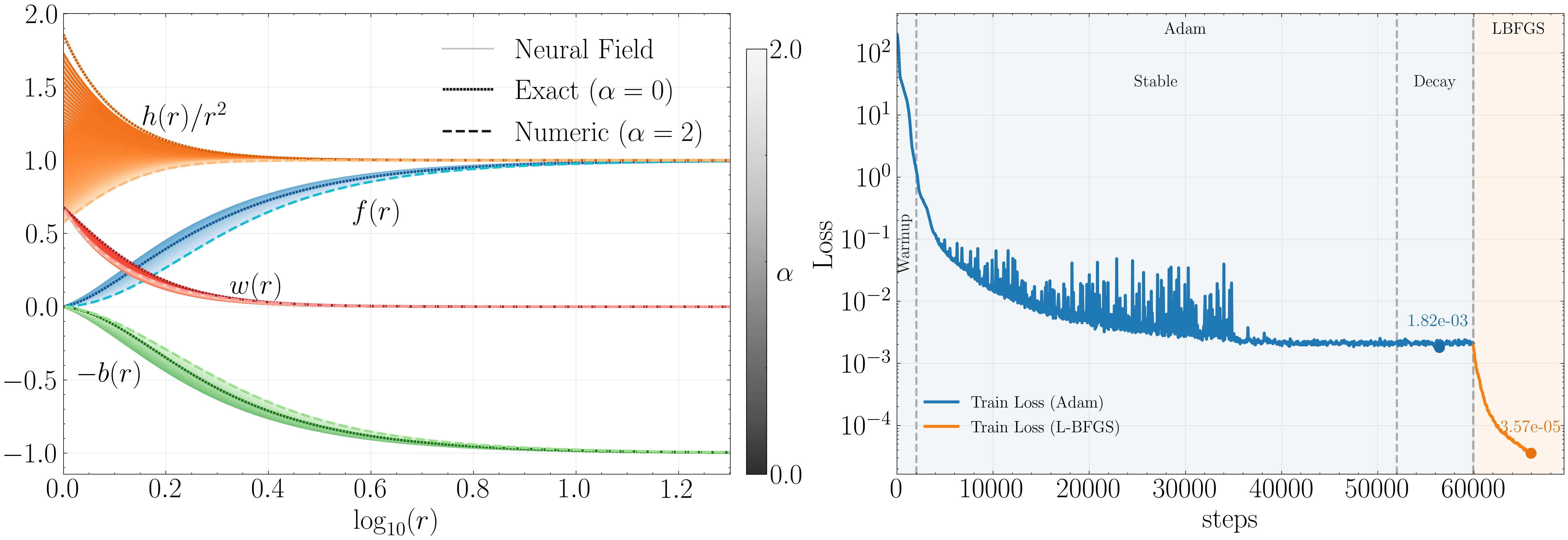}
    \caption{EGB 5D parametric solution for $\Omega_H = 0.68$ with $\alpha \in [0, 2.0]$ and the corresponding loss function.}
    \label{fig:EGB_5D_Oh_0.68}
\end{figure}

\begin{figure}[h!]
    \centering
    \includegraphics[width=1.0\linewidth]{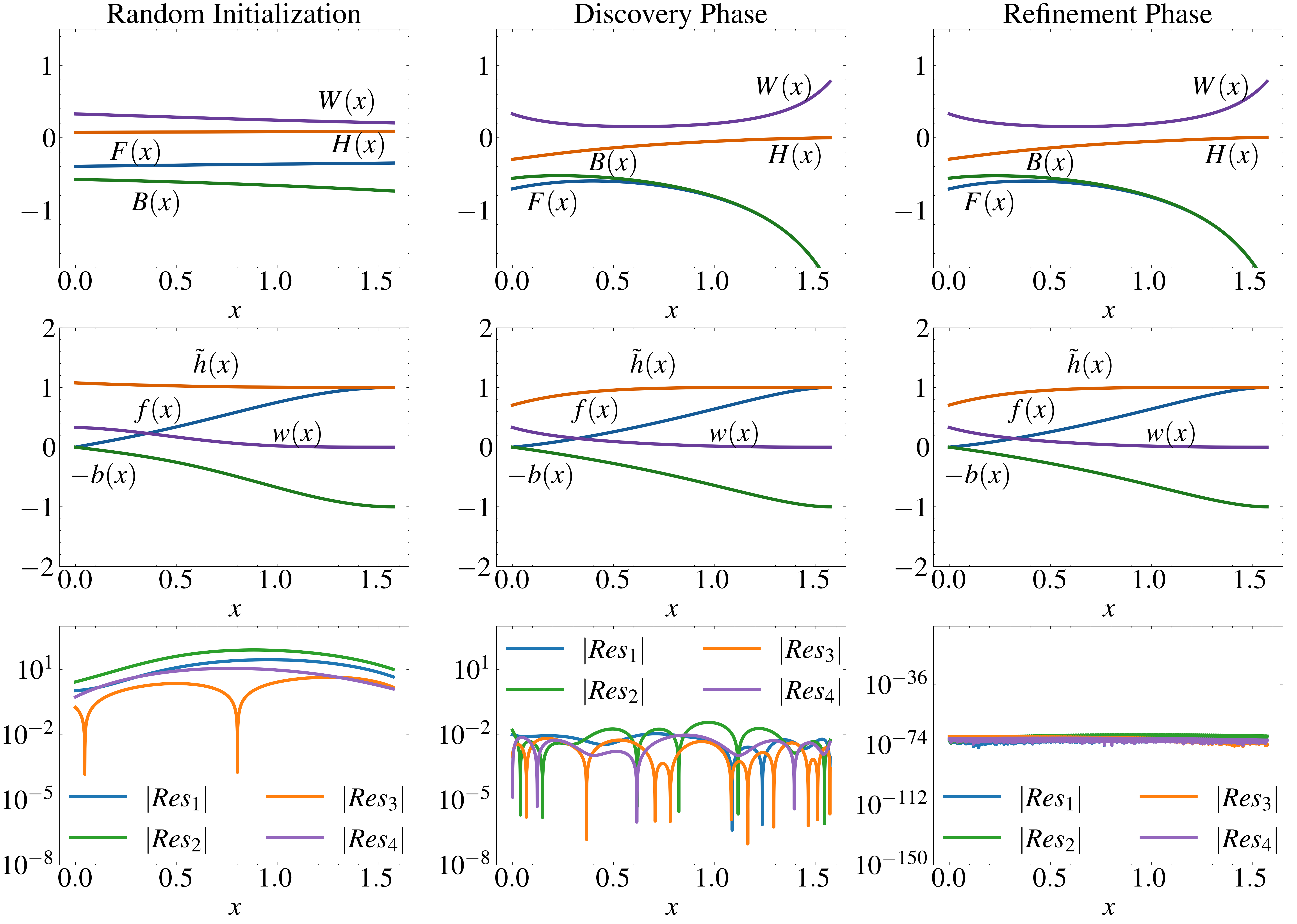}
    \caption{Row 1: upper-case functions of $x$. Row 2: lower-case functions of $x$. Row 3: equation residuals. Each column compares one of the three phases of the method for the Einstein-Gauss-Bonnet solution with $\Omega_H = 0.33$ and $\alpha = 3.0$. Phase 0 shows the network output after random initialization (Random Initialization), Phase 1 the output after the neural-network \textit{L-BFGS} optimization (Discovery Phase), and Phase 2 the output after the Spectral Solver optimization (Refinement Phase).}
    \label{fig:EGB_phases_solution}
\end{figure}


\begin{figure}[h!]
    \centering
    \includegraphics[width=1.0\linewidth]{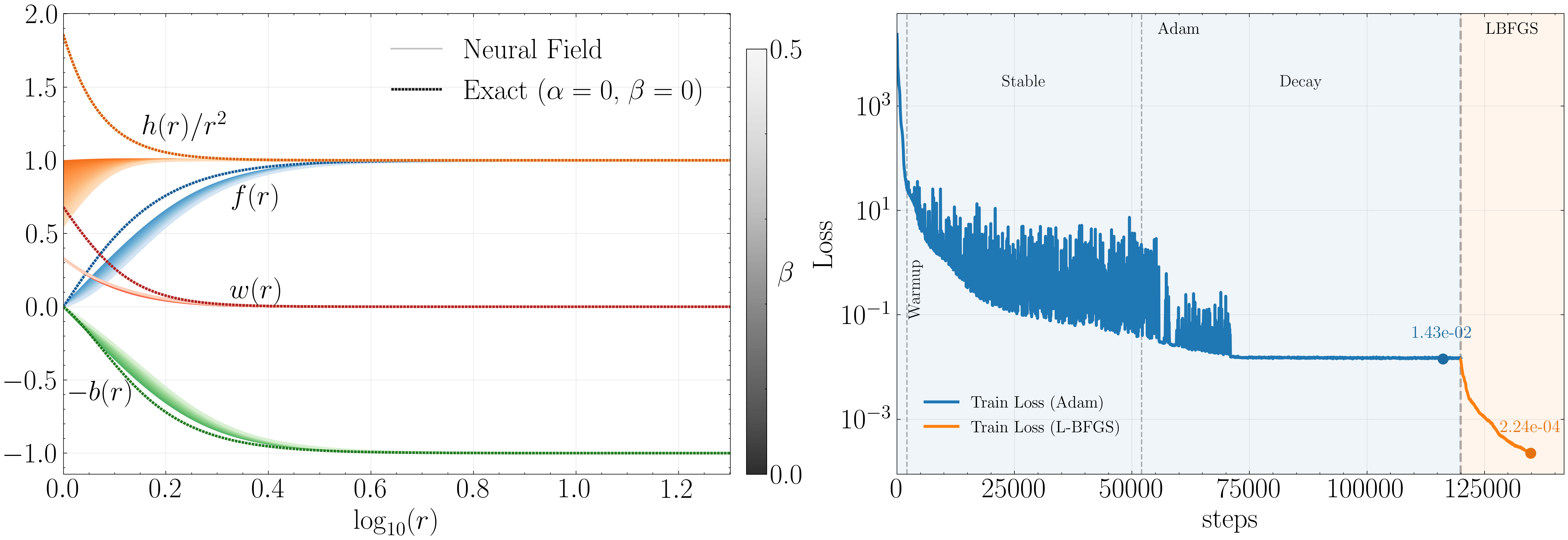}
    \caption{Lovelock 7D parametric solution for $\Omega_H = 0.33$ with $\alpha=0.5$ and $\beta \in [0, 0.5]$ presented in the main text and the corresponding loss function.}
    \label{fig:loss_Lovelock_7D}
\end{figure}

\begin{figure}[h!]
    \centering
    \includegraphics[width=1.0\linewidth]{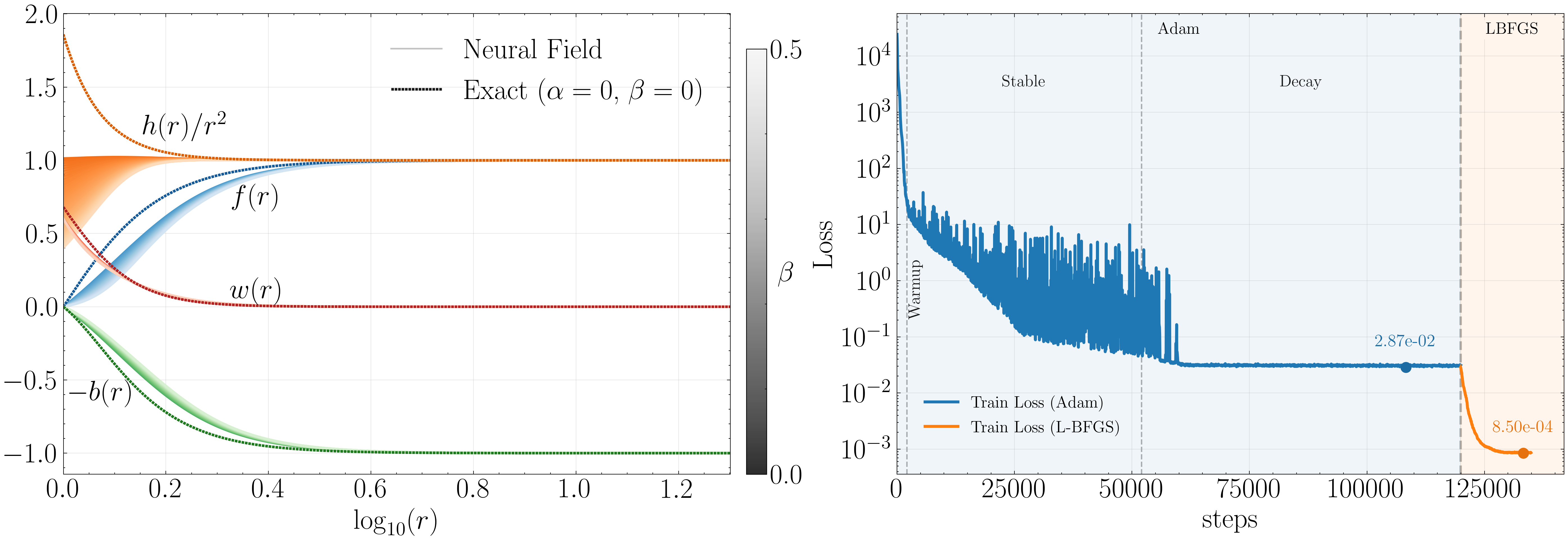}
    \caption{Lovelock 7D parametric solution for $\Omega_H = 0.6$ with $\alpha =0.5$ and $\beta \in [0, 0.5]$ and the corresponding loss function.}
    \label{fig:lovelock_7D_Oh_0.6}
\end{figure}

\begin{figure}[h!]
    \centering
    \includegraphics[width=1.0\linewidth]{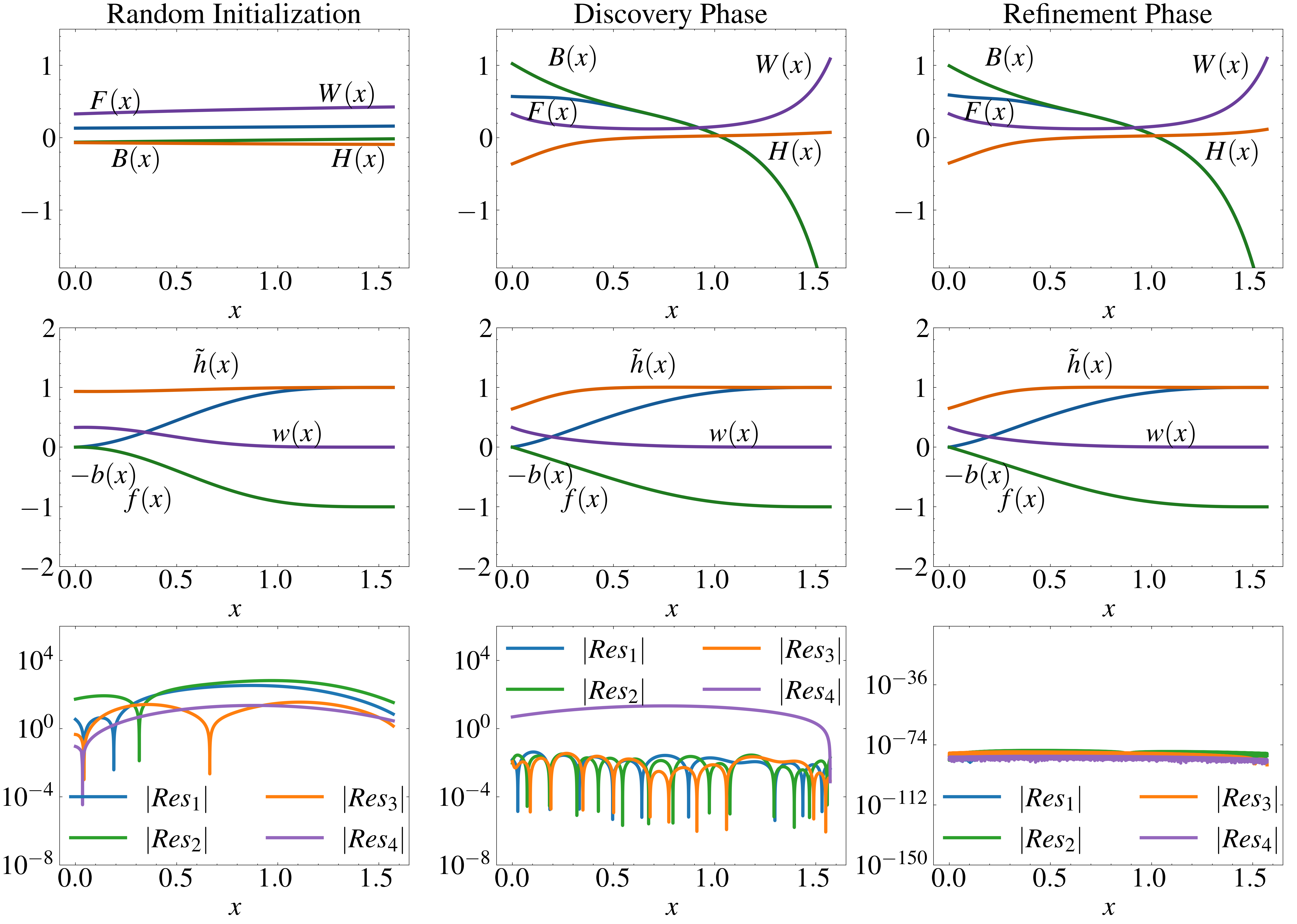}
    \caption{Row 1: upper-case functions of $x$. Row 2: lower-case functions of $x$. Row 3: equation residuals. Each column compares one of the three phases of the method for the cubic Lovelock solution with $\Omega_H = 0.33$, $\alpha = 0.5$, and $\beta = 0.4$. Phase 0 shows the network output after random initialization (Random Initialization), Phase 1 the output after the neural-network \textit{L-BFGS} optimization (Discovery Phase), and Phase 2 the output after the Spectral Solver optimization (Refinement Phase) using $N=500$ collocation points.}
    \label{fig:Lovelock_Phases_solution}
\end{figure}

\clearpage

\begin{table}[h!]
\centering
\setlength{\tabcolsep}{8pt}
\renewcommand{\arraystretch}{1.0}
\begin{tabular*}{0.95\linewidth}{@{\extracolsep{\fill}}|l|l|c|}
\hline
\multicolumn{3}{|c|}{\textbf{Configuration 5D EGB rotating solution}} \\
\hline \hline
\textbf{Category}  & \textbf{Description} & \textbf{Value} \\ 
\hline
Networks & Activation Function & \textit{Swish} \\ 
 & Architecture (Four FCNNs) & [2, 20, 20, 20, 1] \\ \hline
 Domain &  Chebyshev distribution in $x$ & $[0, \pi/2]$ \\
  &  Uniform distribution in $\alpha$ & $[0, 3]$ / $[0, 2]$  \\ \hline
 Physical Quantities  & $r_H$ fixed value & $1.0$ \\
  & $\Omega_H$ Strong BC & $0.33$ / $0.68$ \\
\hline \hline
 \multicolumn{3}{|c|}{\textbf{Discovery phase (Neural Field)}}  \\
 \hline 
\multicolumn{3}{|c|}{\textbf{Stage 1 Global Exploration:} Adam Optimizer and Scheduler} \\
\hline
 & Total Steps & $60\,000$ \\
 & Collocation Points ($N_1$) & $6\,000$ \\
 & Initial Learning Rate & $1.0 \times 10^{-6}$ \\
 & Stable Learning Rate & $1.0 \times 10^{-3}$ \\
 & Final Learning Rate & $1.0 \times 10^{-7}$ \\
\hline
\textbf{Scheduler} & Type & Warmup-Stable-Plateau-Cosine \\
 & Warmup Steps & $2\,000$ \\
 & Stable Steps & $30\,000$ \\
 & Plateau Patience & $1\,000$ (Factor $0.5$) \\
 & Reductions before Cosine & $3$ \\
 & Cosine Steps & $3\,000$ \\
\hline
 \multicolumn{3}{|c|}{\textbf{Stage 2 Local Optimization:} L-BFGS Optimizer} \\
\hline
\textbf{Optimization} & Total Steps & $5\,000$ \\
 & Collocation Points ($N_2$) & $16\,000$ \\
 & Learning Rate & $1.0$ \\
 & Max Iterations & $10$ \\
 & History Size & $50$ \\
 & Line Search & Strong Wolfe \\
 & Tolerances (Grad / Change) & $10^{-8}$ / $10^{-10}$ \\
\hline 
 \end{tabular*}
 
\vspace{0.03cm}
 
 \begin{tabular*}{0.95\linewidth}{@{\extracolsep{\fill}}|l|l|c|c|}
\hline
\multicolumn{4}{|c|}{\textbf{Refinement phase (Pseudospectral Gauss--Newton Optimization)}} \\
\hline

    & Working Precision
    & \multicolumn{2}{c|}{$300$ digits} \\
\hline
\textbf{Category}
    & \textbf{Stage}
    & $\boldsymbol{N}$
    & $\boldsymbol{\lambda}$ \\
\hline
\textbf{Continuation}
    & Stage 1 -- Part 1
    & $20$ & $10^{-2}$ \\
    & Stage 1 -- Part 2
    & $20$ & $1$ \\
    & Stage 2
    & $50$ & $1$ \\
    & Stage 3
    & $100$ & $1$ \\
    & Stage 4
    & $200$ & $1$ \\
\hline
\end{tabular*}
\caption{Specific configuration of the rotating black hole solution with equal angular momenta for the 5D EGB theory.}
\label{tab:EGB5D solution}

\end{table}

\begin{table}[h!]
\centering
\setlength{\tabcolsep}{8pt}
\renewcommand{\arraystretch}{1.0}
\begin{tabular*}{0.95\linewidth}{@{\extracolsep{\fill}}|l|l|c|}
\hline
\multicolumn{3}{|c|}{\textbf{Configuration 7D Lovelock rotating solution}} \\
\hline \hline
\textbf{Category} & \textbf{Description} & \textbf{Value} \\ 
\hline
Networks
    & Activation Function
    & \textit{Swish} \\ 
    & Architecture (Four FCNNs)
    & [2, 64, 64, 64, 64, 1] \\
\hline
Domain
    & Chebyshev distribution in $x$
    & $[0,\pi/2]$ \\
    & Uniform distribution in $\beta$
    & $[0,0.5]$ \\
\hline
Physical Quantities
    & $r_H$ fixed value
    & $1.0$ \\
    & $\alpha$ fixed value
    & $0.5$ \\
    & $\Omega_H$ Strong BC
    & $0.33$ / $0.6$ \\
\hline \hline
\multicolumn{3}{|c|}{\textbf{Discovery phase (Neural Field)}} \\
\hline
\multicolumn{3}{|c|}{\textbf{Stage 1 Global Exploration:} Adam Optimizer and Scheduler} \\
\hline
    & Total Steps
    & $120\,000$ \\
    & Collocation Points ($N_1$)
    & $6\,000$ \\
    & Initial Learning Rate
    & $1.0 \times 10^{-6}$ \\
    & Stable Learning Rate
    & $1.0 \times 10^{-3}$ \\
    & Final Learning Rate
    & $1.0 \times 10^{-7}$ \\
\hline
\textbf{Scheduler}
    & Type
    & Warmup-Stable-Plateau-Cosine \\
    & Warmup Steps
    & $2\,000$ \\
    & Stable Steps
    & $50\,000$ \\
    & Plateau Patience
    & $1\,000$ (Factor $0.5$) \\
    & Reductions before Cosine
    & $3$ \\
    & Cosine Steps
    & $3\,000$ \\
\hline
\multicolumn{3}{|c|}{\textbf{Stage 2 Local Optimization:} L-BFGS Optimizer} \\
\hline
\textbf{Optimization}
    & Total Steps
    & $15\,000$ \\
    & Collocation Points ($N_2$)
    & $12\,000$ \\
    & Learning Rate
    & $1.0$ \\
    & Max Iterations
    & $10$ \\
    & History Size
    & $50$ \\
    & Line Search
    & Strong Wolfe \\
    & Tolerances (Grad / Change)
    & $10^{-8}$ / $10^{-10}$ \\
\hline

\end{tabular*}

\vspace{0.03cm}
 
 \begin{tabular*}{0.95\linewidth}{@{\extracolsep{\fill}}|l|l|c|c|}
\hline
\multicolumn{4}{|c|}{\textbf{Refinement phase (Pseudospectral Gauss--Newton Optimization)}} \\
\hline

    & Working Precision
    & \multicolumn{2}{c|}{$300$ digits} \\
\hline
\textbf{Category}
    & \textbf{Stage}
    & $\boldsymbol{N}$
    & $\boldsymbol{\lambda}$ \\
\hline
\textbf{Continuation}
    & Stage 1 -- Part 1
    & $20$ & $10^{-2}$\\
    & Stage 1 -- Part 2
    & $20$ & $1$  \\
    & Stage 2
    & $50$ & $1$  \\
    & Stage 3
    & $100$ & $1$  \\
    & Stage 4
    & $200$ & $1$  \\
\hline
\end{tabular*}

\caption{Specific configuration of the rotating black hole solution with equal angular momenta for the 7D Lovelock theory.}
\label{tab:7Dsolution}
\end{table}

\end{document}